\documentclass{article}
\font\cero=cmss10 scaled 1728 
\usepackage[utf8]{inputenc} 
\usepackage{multirow}\usepackage{tensor}
\usepackage{verbatim}
\usepackage{amsfonts}
\usepackage{graphicx}
\usepackage{amssymb}
\newcommand{\bra}{\left\langle}
\newcommand{\ket}{\right\rangle}
\usepackage{float}
\usepackage{comment}
\excludecomment{toexclude}
\begin{document}
\begin{flushleft}
{\cero  Strong/weak duality symmetries for Jacobi--Gordon field theory through elliptic functions}\\
\end{flushleft} 
{\sf R. Cartas-Fuentevilla, K. Peralta-Martinez, and D. A.  Zarate-Herrada}\\
{\it Instituto de F\'{\i}sica, Universidad Aut\'onoma de Puebla,
Apartado postal J-48 72570, Puebla, Pue. , M\'exico}; 

{\sf J.L.A. Calvario-Acocal}\\
{\it Escuela de Ciencias, Universidad Aut\'onoma Benito Ju\'arez de Oaxaca,
Apartado postal 68120, Oaxaca de Ju\'arez, Oaxaca, M\'exico}; \\

ABSTRACT:  By using the scheme of Jacobi elliptic functions with their duality symmetries  we present a  formulation of the Jacobi- Gordon field theory that will manifest the strong/weak coupling duality at classical level; for certain continuous limits for the elliptic modulus the model will reduce to the standard sin/sinh Gordon field theories, for which such a strong/weak duality is known only at the level of the S-matrix.
It is shown that the so called self-dual point for the standard sin/sinh Gordon field theory that divides the strong and the weak coupling regimes, corresponds only to one point of a set of fixed points under the duality transformations for the elliptic functions. The potentials constructed in terms of elliptic functions have a critical behavior near that self-dual point, showing a change of topology; in the weak coupling regime the vacuum topology implies that there exists the possibility of formation of topological defects,  and in the strong regime coupling there no exists the possibility of formation of those defects. Furthermore, the equations of motion can be solved in exact form in terms of the inverse elliptic functions; in a case the  kink-like solitons asso\-cia\-ted with the maxima of the potential can decay to  cusp-like solitons associated with the minima. The polynomial expansions of the generalized models show a critical behavior at certain self-dual points; such points define the regions where the spontaneous symmetry breaking scenarios are po\-ssi\-ble.
By invoking the duality symmetries for the elliptic functions, an explicit relation between the original potentials and their dual versions are constructed; with this relationship,
 an approaching to a specific self-dual point is considered for our generalized models.\\

\noindent KEYWORDS: Jacobi elliptic functions, field theories in lower dimensions, nonlinear equations, exact solutions, solitons; duality symmetries; self-dual points.

\section{Introduction}

The sinh-Gordon field theory in $1+1$ dimensions constitute the simplest relativistic model that is integrable,
\begin{eqnarray}
\label{shg} 
	S = \int  dx^2[(\partial_{\mu} \phi)^2 - 2\mu\cosh(b\phi)];
	\label{intro1}
 \end{eqnarray}
 and its spectrum corresponds to a single massive particle.
 Although there is not a clue on  the strong/weak coupling duality, namely $b\rightarrow 1/b$, at the level of the action (\ref{intro1}), the corresponding S-matrix possesses such a duality symmetry \cite{Arin}. Furthermore, the properties of the model in the strong coupling regime  for $b>1$, are obtained by an analytic continuation of the weak coupling regime for $0<b<1$; the special value $b=1$, the so called self-dual point, is dividing both regimes.
 However, since such a duality is not manifest at the level of the classical or at level of the quantum action, this technique based on the analytic manipulation has been severely questioned. 
A categorical not as answer was given recently in \cite{denis} on that analytical manipulation beyond the self-dual point, by proposing that in that regime a background  charge must be introduced; in this manner, the model has in this regime massless renormalization group flows between two different conformal field theories, and describes thus a massless phase, as opposed to the $b<1$ regime associated with a massive particle, and with a zero background charge; in this construction, the massive phase can be defined as a perturbation of a free massless boson in the ultraviolet. The proposal in \cite{denis} is robust since it reproduces correctly the freezing transitions in the multi-fractal exponents for  a Dirac fermion in a random magnetic field.

In \cite{konik} the truncated spectrum methods were developed for studying the sinh-Gordon model at finite volume; however, by varying the coupling constant near the self-dual point, the approach breaks down; for addressing these limitations, different strategies are explored, and the preliminary results show that the theory may describe a massless phase beyond the self-dual point; however, various aspects of such a phase are unknown, and its properties can be very different from those that one can infer from the structure of its S-matrix.

The main purpose of the work at hand is, by using an elliptic functions based formulation, to provide a classical formulation that will manifest explicitly such a duality through certain set of transformations connecting those functions. The key role is played by the elliptic modulus, which allow us to connect different regimes for a generalized model; in particular, at certain limits for the modulus, the model will reduce to the sinh-Gordon model, and the sin-Gordon model respectively.

 In general there exist three dualities, namely, the Jacobi imaginary transformation, Jacobi ima\-ginary modulus transformation, and the Jacobi real transformation. These dualities allow to define the elliptic functions on the full real line for the square modulus ($\kappa$), with certain special points that define six sub-intervals,
\begin{eqnarray}
\kappa=(-\infty,-1)\cup (-1,0)\cup (0,1/2)\cup(1/2,1)\cup(1,2)\cup(2,+\infty);
\label{full}
\end{eqnarray}
the special points define the different trigonometries associated with those functions, namely, for $\kappa=0$, the curve is a circle, for $0<\kappa<1$ the curve is an ellipse, for $\kappa>1$ the curve is a hyperbola, which degenerates to two lines for $\kappa=1$; moreover, the values $\kappa=(-1,1/2,2)$ correspond to three versions of the lemniscate trigonometric functions in terms of elliptic functions; in fact, such representations are connected to each other by the dualities. Additionally each special value in the above expression corresponds to a self-point for certain duality transformation between elliptic functions; hence, one has this geometrical perspective for facing the problem of approaching to the self-dual points of any elliptic functions based formulation in field theory; for example, in \cite{konik} it is recognized that 
the problem with the analytic continuation for values of $b>1$ there exists in fact from the values for $b>1/\sqrt{2}$. The answer is clear from the perspective based on the dualities of the elliptic functions, since the value $\sqrt{\kappa}=1/\sqrt{2}$ corresponds also to a self-dual point. In fact, one can consider that there exists a fundamental interval in the above expression (\ref{full}), namely
$0\leq\kappa\leq 1/2$, and then the full real line is covered by invoking the duality transformations. These two self-dual points, $(1/2,1)$, will play a crucial role in the approach at hand.

Due to the duality transformations between  the elliptic functions, the possible functional forms for the square modulus are, $\kappa$,  $1/\kappa$, $1-\kappa$, $\kappa/(\kappa-1)$, $1-1/\kappa$, and $1/(1-\kappa)$; thus, considering that the original modulus is $\kappa$, the special points in the expression (\ref{full}) can be determined as self-dual points by matching the original modulus with the second, third, and fourth different forms, namely, $\kappa=1/\kappa$, {\it i.e} $\kappa=\pm1$, $\kappa=1-\kappa$, {\it i.e} $\kappa=1/2$, $\kappa=\kappa/(\kappa-1)$, {\it i.e} $\kappa=(0,2)$; these self-dual points are real,  as opposed to the last two forms that will lead to complex self-dual points;  these forms have the same two self-dual points on the complex plane, namely $\frac{1}{2}(1\pm i\sqrt{3})$.
Therefore, the self-dual points 
can be viewed as fixed points under duality transformations. For example, we can assume the fundamental interval $0\leq\kappa\leq 1/2$ as starting point; the transformation $\kappa\rightarrow1-k $ will transform that interval into $1/2\leq\kappa\leq 1$; now, the action of the transformation $\kappa\rightarrow 1/\kappa$ on the later interval will lead to the interval $1\leq\kappa\leq 2$, and so on; note that in these examples the self-dual points correspond to fixed points under the transformations.
Furthermore, the six functional forms for the modulus described above can be constructed as invariant forms under the action of the modular group; this group theory point of view was developed by F. Klein
\cite{klein} at the end of the nineteenth century; for a recent review on this approach, and the respective analysis on the solutions for the simple pendulum given in terms of elliptic functions, see \cite{linares}.

 The elliptic functions have appeared in a wide variety of systems in nonlinear physics, and historically 
the classical systems such as the simple pendulum, spherical pendulum, Duffing oscillator, the bead on the hoop (see \cite{hoop}, and \cite{lawden}), etc, were the first problems solved in terms of those functions; in field theory 
several cases are in order, KdV equation, mKdV equation, Boussinesq equation,  nonlinear Klein-Gordon equation, etc; all these examples in field theory have been considered in \cite{zhao}, where a  Jacobi elliptic function expansion approach has been developed. Moreover, the solutions in terms of elliptic functions for a quartic massless field theory lead to a dispersion relation for a massive wave, and for a quartic scalar field theory, one obtains a renormalized mass through a term depending on the self-coupling parameter \cite{frasca}. However, the duality symmetries for these nonlinear system in field theory have not been explored, and their possible physical meaning is not well known. In this work we  
formulate a generalized Jacobi-Gordon field theory as a generalization of sin and sinh-Gordon field theories and we study its solutions, and its duality symmetries; in particular we attempt an approaching to the self-dual point, from the classical formulation based on the elliptic functions.

As an antecedent we describe briefly the physical meaning of the duality symmetries for the classical system better understood nowadays, the simple pendulum; the role of the duality is well known, it connects an oscillation of small amplitude (associated with a small modulus), with an oscillation of large amplitude (associated with a large modulus). Specifically the nonlinear equation for the simple pendulum, $\ddot{\theta}+sin\theta=0$, and the conservation of the energy, $\frac{E}{2}=\frac{\dot\theta^2}{4}+sin^2\frac{\theta}{2}$, are transformed into  $\ddot{\theta'}+sin\theta'=0$, and $1-\frac{E}{2}=\frac{\dot\theta'^2}{4}+sin^2\frac{\theta'}{2}$, under the transformation $t'\rightarrow it$, $\theta'\rightarrow \theta+\pi$, and $\frac{E}{2}\rightarrow1-\frac{E}{2}$; for this system the square modulus corresponds to the energy, $\kappa=\frac{E}{2}$, and thus the complementary modulus is $\kappa'=1-\kappa$, with $0<\kappa<1$, and $0<\kappa'<1$. The limit $\kappa\rightarrow 1$, it means $E\rightarrow 2$, corresponds to the maximum value for the oscillatory motion, which is excluded since it is an unstable equilibrium point; thus, if the energy is less than this value, the pendulum 
will oscillate. In this manner, the above transformation will map a regime for small (large) amplitude, it means small (large) $\kappa$, to a regime for large (small) amplitude, it means  large (small) $\kappa'$;
note that the product of the transformation $\kappa\rightarrow1-\kappa$ corresponds to the third functional form for the modulus discussed previously, for which the self-dual point $\kappa=1/2$ is separating both regimes; for this value the orbit is invariant under the above transformation \cite{rajeev}. This system with its  small/large amplitude symmetry and with a self-dual point separating the two regimes
may to represent the mechanical system analogous to the weak/strong coupling regimes for models such as sin/sinh Gordon field theory.

However there exists another duality symmetry for the simple pendulum that is little known, but it can be established in a direct way by using the symmetries for the elliptic functions, and that it will result closer to that duality symmetry for the field theories under consideration; for the new duality the solutions in terms of elliptic functions will appear explicitly, as opposed to the previous case, and additionally the analytical continuation for time will be not invoked. First, the transformation $\sin\frac{\theta}{2}=\sqrt{\kappa}\sin\varphi$ allows us to identify directly the new variable $\varphi$ with the inverse of the elliptic integral that defines the oscillation period for the pendulum, $t=\int^{\varphi}_{0}(1-\kappa\sin^2\phi)^{-1/2}d\phi$, and thus $\varphi=Am(t,{\kappa})$, where $Am$ stands for the Jacobi amplitude, and the new variable satisfies a dual equation, $\ddot\varphi+\kappa \sin\varphi=0$;  therefore $\sin\varphi=sn(t,{\kappa})$, according to the definition for the elliptic sine function $sn$, which has the duality symmetry $\sqrt{\kappa}sn(z,\kappa)=sn(\sqrt{\kappa}z,1/\kappa)$, it means the global factor depending on the elliptic modulus can be absorbed into the argument and inverting the elliptic modulus. In this manner we have that $\frac{\theta}{2}=Am(\sqrt{\kappa}t,{1/\kappa})$; thus, for small $\kappa$, $\varphi$ describes small oscillations, but it will correspond to large oscillations  for the variable $\theta$, and conversely for large values of $\kappa$. In this duality transformation the modulus transforms as $\kappa\rightarrow 1/\kappa$, and the for self-dual point $\kappa=1$, the orbits for both dynamical variables are the same; note that the self-dual point is separating again the regimes for small and large oscillations. It is just  this kind of duality that the simple pendulum  manifests at classical level, that the sinh-Gordon model manifests at  the level of its S-matrix.

With these antecedents we have seen that the elliptic functions formalism shows certain virtues for separating different regimes for a classical physical system, and it motivates us to construct a formulation for the sin/sihn Gordon field theory that can to manifest, at classical level, the duality symmetry  that is known only at S-matrix level. For this purpose in the next section we propose an elliptic functions based formulation of a $(1+1)$ Lagrangian for a scalar field, that for certain limits of the elliptic modulus the sin/sinh Gordon field theories are obtained as continuous limits.  Actually we shall develop two independent models for our generalized formulation of sin/sihn Gordon theory (sections \ref{fc}, and \ref{sc}); the respective potentials will be identified with certain quotients of the fundamental elliptic functions, which will allow us to solve analytically the equations of motion in two scenarios, namely, the complete theory and the expansions around the maxima/minima for the potentials (sections \ref{cuspsoliton}, and \ref{kinksoliton}). The solutions correspond to different field configurations associated with cusp-like solitons, and kink-like solitons; in a case, the kink-like solitons identified with the maxima of the potentials, may decay to cusp-like solitons identified with the minima. 

The elliptic functions based potentials will show a critical behavior at $\kappa\rightarrow 1$, reveling the role that the self-dual points play at classical level, namely to separate explicitly two regimes;
for $\kappa<1$
they correspond to periodic potentials with an infinite collection of vacua, with the possibility of  formation of topological defects (section \ref{quartic}); for $\kappa>1$ the potentials will have a change of topology, and correspond a collection of disconnected vacua, without the possibility of formation of topological defects (section \ref{kappabig}). Moreover, the polynomial expansions of the potentials  around the maxima and minima for $\kappa<1$, show an interesting structure; the quartic field theory approach is critical at two self-dual points, at $\kappa=1$, for which the complete theory is also critical, and at a second self-dual point, $\kappa=1/2$;
in fact, the spontaneous symmetry breaking scenarios can be constructed only within the fundamental 
interval $0<\kappa<1/2$ (section \ref{quartic}). In section \ref{conlimit}  the sin-Gordon and sinh-Gordon models are obtained as the continuous limits for $\kappa\rightarrow 0$, and $\kappa\rightarrow 1$; specifically the double versions that involve the fundamental harmonics and higher harmonics are constructed from our generalized models. In section \ref{GLS} the Jacobi duality symmetries for the elliptic functions based potentials are established as Lagrangian symmetries; additionally the duality transformations generate singularities on the potentials, and they are determined by the integral periods. Specifically we shall focus on duality transformations 
that have as self-dual point the value $\kappa=1$ for the modulus, closer to the duality symmetry identified for the sin/sinh-Gordon models at quantum level. The approaching to the self-dual point is analyzed by expanding the potentials around the value $\kappa=1$, and analyzing the limits for $\kappa<1$, and for $\kappa>1$.

\section{Lagrangian and the equation of motion}
For the Lagrangian of the general form
\begin{eqnarray}
L(\phi)=\partial_{\mu}\phi\cdot\partial^{\mu}\phi-V(\phi),
\label{lag}
\end{eqnarray}
the corresponding equation of motion reads
\begin{eqnarray}
\frac{1}{c^2}\frac{\partial^2 \phi}{\partial t^2}-\frac{\partial^2 \phi}{\partial x^2}=-\frac{dV(\phi)}{d\phi},
\label{em}
\end{eqnarray}
where we have considered by simplicity a $1+1$ background; 
by introducing the wave variable $\xi \equiv k_{0}(x-c_{0}t)$, where $c_{0}$ represents the wave  velocity, the above equation will read
\begin{eqnarray}
k_{0}^2(1-\frac{{c_{0}}^ 2}{c^2})\frac{d^2\phi}{d\xi^2}=\frac{dV}{d\phi};
\label{soliton}
\end{eqnarray}
note that if the wave velocity equals to the light speed, $c_{0}=c$, then $V$ is a constant, and we have a free wave propagating to the light speed. 
Hence, we consider that $c_{0}<c$, and $(1-\frac{{c_{0}} ^2}{c^2})>0$; therefore, the Eq. (\ref{soliton}) can be rewritten in the standard form,
\begin{eqnarray}
\pm (x-c_{0}t)=\sqrt{\frac{1-\frac{{c_{0}} ^2}{c^2}}{2}}\int \frac{d\phi}{\sqrt{V(\phi)}}.
\label{soliton1}
\end{eqnarray}
From this point we shall use the natural units with $\hbar=1=c$.
In order to face the construction of solutions for this equation of motion, and to construct a field theory with duality symmetries at classical level,
we shall consider the following exact integral involving elliptic functions
\begin{eqnarray}
\int \frac{dn(z)}{sn(z)} dz =\frac{1}{2} \ln \frac{1-cn(z)}{1+cn(z)}, \label{ei1}\\ \quad
\int \frac{dn(z)}{cn(z)} dz =\frac{1}{2} \ln \frac{1+sn(z)}{1-sn(z)};
\label{ei2}
\end{eqnarray}
where $sn$, $cn$, and $dn$ stand for elliptic sine, elliptic cosine, and delta function respectively; our strategy consists in
 identifying appropriately the potential $V(\phi)$ with the integrants in the above expressions. Hence, we shall be able of constructing field configurations with a potential that will reduce to the potentials for Sine-Gordon, and Sinh-Gordon field theories in certain limits for the elliptic modulus. In particular, in the quartic order field theory approximation one obtains an infinite family of theories with SSB scenarios labelled by the elliptic modulus.
 
 The quotients of elliptic functions as integrants in the expressions (\ref{ei1}), and (\ref{ei2}), that will define our potentials and that play the key role in the approach at hand, have appeared in several geome\-trical scenarios as fundamental mappings; 
for example, the expressions 
$x=sn(\phi,\kappa)/dn(\phi,\kappa)$, and $y=cn(\phi,\kappa)/dn(\phi,\kappa)$ can be considered as the generalization for the coordinates of the points in the unit circle, namely $x=sin(\phi)$, and $y=cos(\phi)$, to the coordinates of the points on the unit ellipse. Furthermore,
for $\phi=\sqrt{2}z$, and $\kappa=1/\sqrt{2}$, the map $w= sn(\sqrt{2}z,1/\sqrt{2})/dn(\sqrt{2}z,1/\sqrt{2})$, defines a conformal transformation from a square to the unit disk in the complex plane; $z$ and $w$ represent the locations in the unit disk and in the square respectively.
\\

\section{First case}
\label{fc}
We first identify the integrant in the Eq. (\ref{soliton1}) with the integrant in the Eq. (\ref{ei2}), namely, $ \frac{dn( \phi ; \kappa)}{cn( \phi; \kappa}=\frac{m}{\sqrt{V_{1}(\phi)}}$, where $m$ is a real parameter with mass units. Additionally the right hand side of the equality (\ref{ei2})
will be used in section (\ref{sntanh}) for constructing explicitly the solution for the field $\phi$ as a function on $(x,t)$.
Thus the potential will read
\begin{eqnarray}
V_{1}(\phi;\kappa)=m^2\frac{cn^2(\phi ; \kappa)}{dn^2(\phi ; \kappa)};\quad
V_{1}(\phi, \kappa=0)=m^2(\cos 2 \phi +1),\quad
V_{1}(\phi, \kappa=1)=m^2;
\label{potlimit}
\end{eqnarray}
hence, for these limiting values of $\kappa$ one obtains the Sine-Gordon potential, and the trivial case for a constant potential; thus, for the elliptic modulus $\kappa\in[0,1]$, the above potential can be considered as a generalization or a deformation of the standard 
Sine-Gordon potential; the figure 1 illustrates this periodic potential for certain values of $m$, and with $\kappa<1$.
We recall that the delta function $dn$ has no zeros, and then the potential is well defined is this interval; note that the potential has the $Z_{2}$ symmetry $\phi\rightarrow -\phi$; the potential has additional symmetries, namely, the first one is inherited from the periodicity of the elliptic functions, and it will be described below; the second one corresponds to the duality symmetry, which is of a subtler type, and it will be described in detail in the section \ref{GLS}.

\begin{figure}[H]
  \begin{center}
   \includegraphics[width=.65\textwidth]{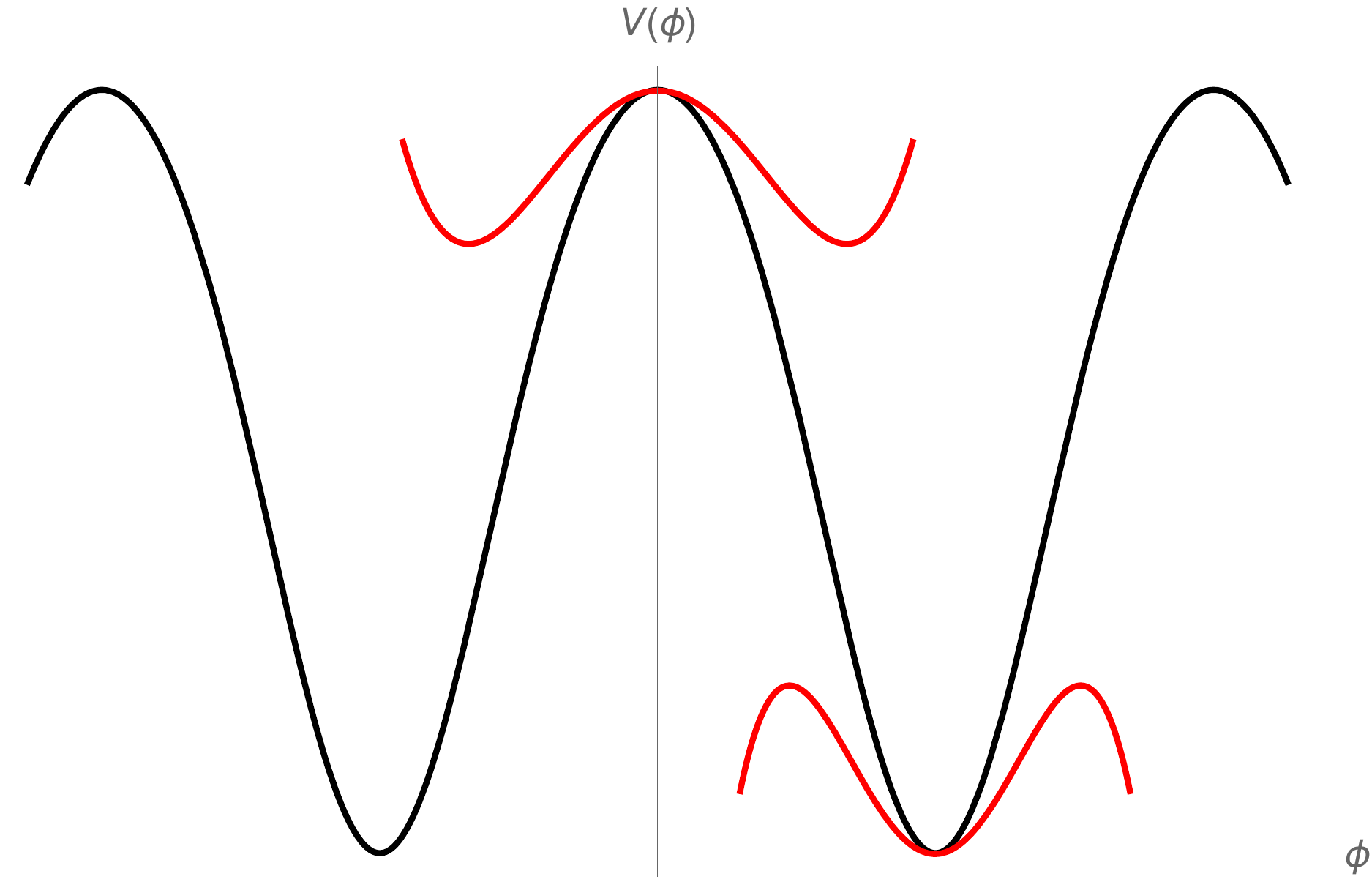}
  \caption{The periodic potential (\ref{potlimit}) in the interval $ \kappa <1$. In the  sub-interval $\kappa<1/2$, a SSB scenario is possible at the maxima. At minima  an up-side down SSB scenario can be constructed in the range $-1< \kappa <1$.}  
  \label{cn}
\end{center}
\end{figure}

Taking into account the second derivative criteria we determine the maxima and minima,
\begin{eqnarray}
    \frac{dV_{1}(\phi,\kappa)}{d\phi}={2 (\kappa-1) m^2} \frac{cn(\phi,\kappa) sn (\phi,\kappa)}{  dn (\phi,\kappa)^3};\label{der}\\
     \frac{d^{2}V_{1}(\phi,\kappa)}{d\phi^{2}} =  2 (\kappa-1) m^2\frac{ -\kappa cn (\phi,\kappa)^4+2 cn(\phi,\kappa)^2+\kappa-1}{dn(\phi,\kappa)^4}; 
     \end{eqnarray}
hence,  $ \frac{dV_{1}(\phi,\kappa)}{d\phi}=0$ only if $sn (\phi,\kappa) = 0$ or $cn (\phi,\kappa) = 0$; the zeros for the elliptic functions, for arbitrary $\kappa$, read
\begin{eqnarray*}
cn(2nK(\kappa), \kappa) = 1; \quad cn((2n+1)K(\kappa), \kappa) = 0;\\
sn (2nK(\kappa), \kappa) = 0; \quad sn ((2n+1)K(\kappa), \kappa) = 1;\\
dn(2nK(\kappa), \kappa)= 1; \quad dn((2n+1)K(\kappa), \kappa) = \sqrt{1-\kappa}; & \quad \textrm{for $n\in \mathbb{Z}$};
\end{eqnarray*}
where
$K(\kappa) = \int_{0}^{\pi/2}\frac{d\theta}{\sqrt{1-\kappa\sin^{2}{\theta}}}$, is the complete elliptic integral. It follows that,
\begin{eqnarray}
      \phi_{_{critical}} = \left\{
    \begin{array}{cc}
        \phi_{_M}\equiv 2nK(\kappa),  & \textrm{Maxima};\\  <\phi> \equiv(2n+1)K(\kappa), & \textrm{Minima};
    \end{array}
\right.  \label{maxmin}
\end{eqnarray}
from these expressions we realize that the (integral) period of the potential will be $2K(k)$. The vacuum configuration is not unique,  instead our potential has multiple vacua, labeled by a discrete parameter $n$ and a continuous elliptic modulus $\kappa$; for example for a vanishing modulus  $K(\kappa=0)=\pi$, corresponding to the sin-Gordon model, the vacua will be labeled by $2\pi n$. The identification of maxima and minima described above, and illustrated in the figure (\ref{cn}), are valid only for the interval $\kappa<1$; for $\kappa>1$ a change of topology in the potential will be present, due to the critical behavior
of the derivatives (\ref{der})  at $\kappa\rightarrow 1$ (see the figure \ref{mmcn}).

Up to this point the only critical point corresponds to the value $\kappa=1$; first, the sign of the derivatives in
Eqs. (\ref{der}) depends on the limit $\kappa\rightarrow 1$ for $\kappa<1$, and for $\kappa>1$ (see the section \ref{kappabig} below); second, the complete elliptic integral diverges at the self-dual point, $lim_{\kappa\rightarrow 1^-} K(\kappa)=\infty$, it means for $\kappa<1$; hence, the expressions (\ref{maxmin}) diverge in the weak coupling regime. For $\kappa>1$, $K(\kappa)$ is extended onto the complex plane, with a divergent real part at the limit $\kappa\rightarrow 1^+$; in general we have that
\begin{eqnarray}
K(\kappa)=\frac{1}{k}[K(1/\kappa)-iK(\frac{\sqrt{k^2-1}}{\kappa})]; \quad \kappa>1;
\label{Kmas}
\end{eqnarray}
thus,
\begin{eqnarray}
lim_{\kappa\rightarrow 1^+}K(\kappa)=+\infty-i\pi/2; \quad \kappa>1;
\label{Kmas}
\end{eqnarray}
where $lim_{_{\frac{1}{\kappa}\rightarrow 1^{-}}}K(1/\kappa)=+\infty$, and $lim_{{\kappa}\rightarrow 1^+}K(\frac{\sqrt{k^2-1}}{\kappa})=\pi/2$. Therefore, the self-dual point $\kappa=1$ is singular on the real line.

 The polynomial expansion for the potential will show a second critical point, namely, the self-dual point $\kappa=1/2$.
The series expansions around the arbitrary maxima and minima are given by,
\begin{eqnarray}
V_{1}(\phi,\kappa)|_{_{max}}  =
    m^2+(\kappa -1) m^2 \left(\phi- \phi_{_M}\right)^2+\frac{2}{3}   (\kappa -1) (\kappa -\frac{1}{2}) m^2 \left(\phi- \phi_{_M}\right)^4+\ldots;
    \label{expan2}
\end{eqnarray}
\begin{eqnarray}
V_{1}(\phi,\kappa)|_{_{mim}}  = 
    m^2 \left(\phi-<\phi>\right)^2-\frac{1}{3}   (\kappa +1) m^2 \left(\phi-<\phi>\right)^4+\ldots;
    \label{expan3}
\end{eqnarray}

The coefficients for these polynomial expansions are depending on the modulus, and show a critical behavior 
at certain self-dual points. Specifically for the expansion (\ref{expan2}) all the coefficients vanish at $\kappa=1$;
moreover the coefficients for the terms of orders $4,8,12,...,$ in $\phi-\phi_{M}$, vanish also at a second self-dual point, 
$\kappa=1/2$, since they have the form $(\kappa-1)(\kappa-\frac{1}{2})P$, where $P$ is a polynomial in $\kappa$ with roots different to self-dual points. For the expression (\ref{expan3}) all coefficients for the terms of orders $4,8,12,...,$ in $\phi-<\phi>$, will vanish at a different and unique self-dual point, $\kappa=-1$.
Note that only the expansion around the maxima retains the original critical point, $\kappa=1$, and a second critical point emerges, $\kappa=1/2$. The expansion around the minima has no a trace of the original critical point, instead a different self-dual point emerges, $\kappa=-1$.

We describe now the ranges for the modulus in which there will exist SSB scenarios for these polynomial expansions.

\subsection{A quartic field theory with SSB scenario:$ \kappa < \frac{1}{2}$}
\label{quartic}
We analyze by simplicity the expansion (\ref{expan2}) around the maximum $\phi =0$;
\begin{eqnarray}
V_{1}( \phi ; \kappa)={m^2}+m^2(\kappa -1) \phi ^ 2+\frac{2}{3}m^2 (\kappa -1)(\kappa - \frac{1}{2})\phi^4 + \mathcal{O} (\phi ^6)+ ... ;
\label{expan1}
\end{eqnarray}
note that the sign of the expansion coefficients are defined by the elliptic modulus; since the spontaneous symmetry breaking scenarios are achieved with a negative definite quadratic term, and a po\-si\-tive definite quartic term, the modulus $\kappa$ must be restricted to the fundamental interval $0\leq \kappa < \frac{1}{2}$, in this approximation; actually the restriction for obtaining  a local SSB scenario is $\kappa<1/2$, it means including the negative values for the modulus.
In contrast, in the interval $\frac{1}{2} < \kappa <1$, the approximation  (\ref{expan1}) has not minima, rather the point $\phi=0$ will correspond to a global maximum. Moreover, the vacuum expectation values in this quartic approximation reads
\begin{eqnarray}
\bra \phi \ket ^2_{_{0}}=\frac{3}{4(\frac{1}{2}-\kappa)}; \quad \kappa <1/2,
\label{vev}
\end{eqnarray}
hence, the vev reaches the minimum value for $\kappa=0$ within the fundamental interval $0 \leq \kappa <1/2$, corresponding to the expansion of the Sine-Gordon case; the vev diverges at the limit $\kappa \to \frac{1}{2}$, and it vanishes for $\kappa\rightarrow -\infty$. This vev for a double well potential constructed around the maxima, must be compared with the vev at the minima given in the Eq. (\ref{maxmin}), for which the divergence occurs at $\kappa\rightarrow 1$, due to the property $lim_{\kappa\rightarrow 1} K(\kappa)=\infty$; hence, the vev diverges for the $\phi^4$-approach  at the first self-dual point, and it diverges for the complete approach at the second self-dual point.
Additionally the mass generated by expanding the potential around $\bra \phi \ket_{_{0}}$ given in Eq. (\ref{vev}), namely, $V(\phi+ \bra \phi \ket_{_{0}}, \kappa)$, will take the form
\begin{eqnarray}
m_{\phi}^2=-2(\kappa-1)m^2, \quad  \kappa <1/2,
\label{genmass}
\end{eqnarray}
which shows, as usual, the duplication and the change of sign respect to the mass term in the expression (\ref{expan1}), within the range $ \kappa <1/2$, previously fixed for  a SSB scenario.

At the self-dual point $\kappa=1/2$, one must to return to the expansion (\ref{expan1}) in order to consider the next sixth order term, since the quartic order term vanishes at $\kappa \to 1/2$. The corresponding coefficient for such higher order term reads $C_{6}(\kappa)\equiv (\kappa-1)[17\kappa(\kappa-1)+2]$, which is not vanishing at $\kappa=1/2$, rather $C_{6} (1/2)>0$, defining hence a possible SSB scenario; see for example \cite{roman} for a numerical description of kink-antikink collisions for a $\phi^6$ model. In fact, the value $\kappa=1/2$ corresponds to the arithmetic mean in the interval where $C_{6}(\kappa)$ is positive definite, namely $\frac{1}{2}-\frac{3}{2\sqrt{17}}<\kappa < \frac{1}{2}+\frac{3}{2\sqrt{17}}$.

Note that the restriction $ \kappa <1/2$ on the elliptic modulus is imposed only in the polynomial approximation at $\phi=0$,
in order to construct a standard SSB scenario;
 in contrast, in the interval $\frac{1}{2} < \kappa <1$, the approximation  (\ref{expan1}) has not minima, rather the point $\phi=0$ will correspond to a global maximum.
Furthermore, for the full expression (\ref{potlimit}) the point $\phi=0$ corresponds a maximum with minima at the both sides for arbitrary values of the modulus within the full range $ \kappa <1$; thus, we have a SSB scenario for the full large field amplitude beyond the quartic field theory approach.
Moreover, due to the periodicity of the potential, we have a full SSB scenario for each maximum for the complete theory.

\subsection{Expanding around an arbitrary minimum: a cusp-like soliton}
\label{cuspsoliton}
We can expand the potential around an arbitrary minimum by using the addition formulae for the elliptic functions, obtaining a secondary functional form
\begin{eqnarray}
V_{1}(min+\Delta\phi; \kappa)=m^2 sn^2(\Delta\phi;\kappa)\approx m^2(\Delta \phi^2-\frac{\kappa+1}{3}\Delta\phi^4+\ldots);
\label{minex}
\end{eqnarray}
where $min$ stands for an arbitrary minimum described in (\ref{maxmin}), and $\Delta\phi$ defines the field fluctuations around such a minimum. Note that this expansion gives another functional form for the potential, namely, $sn^2$, which is expanded in a  polynomial for obtaining locally the standard polynomial SSB scenarios; this polynomial expansion is consistent with the expansion (\ref{expan3}) by identifying $\Delta \phi=\phi-<\phi>$. In this polynomial approach, the mass term is positive definite and independent on the modulus, and additionally the quartic order term is negative definite in the interval $-1< \kappa <1$; thus we have the up-side down SSB scenario described at a minimum in the figure \ref{cn}.
Note that the sign of the mass term corresponds to that generated by SSB in the standard $\lambda\phi^4$ model; 
hence, a standard SSB mechanism will work in the usual form from the dome at the maximum, into the vacuum at the minimum. In the interval $\kappa<-1$ the quartic field theory around the minima will have an absolute minimum.

\begin{figure}[H]
  \begin{center}
   \includegraphics[width=.55\textwidth]{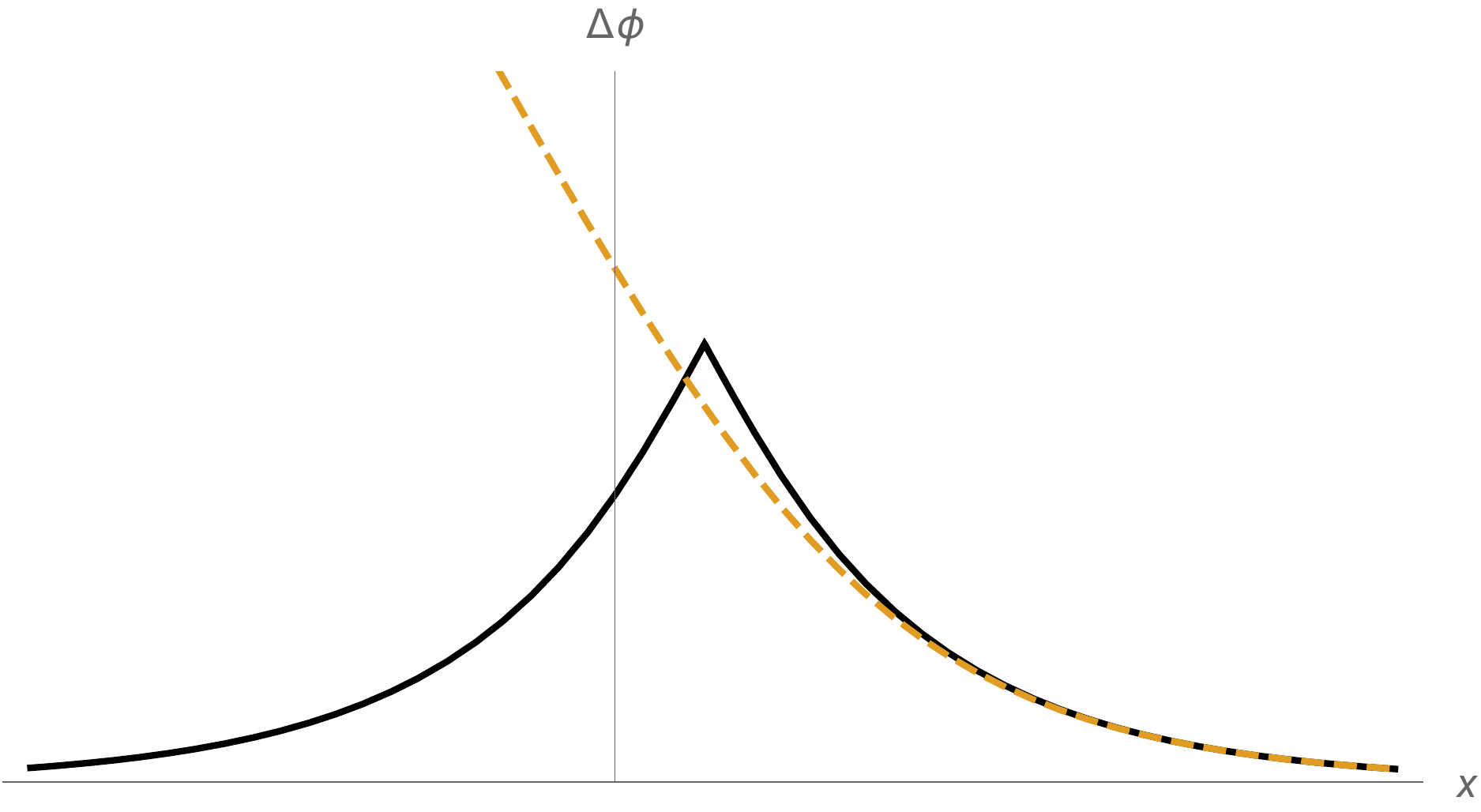}
  \caption{ A cusp-like profile as the configuration of the soliton (\ref{am3}) for $t=0$; the black curve represents the profile for $\kappa<1$;  the dashed curve represents the profile for $\kappa=1$; in both cases the sign $+$ in the argument has been chosen;
  the choice $-$ in the argument will reflect the profiles on the $x$-axis according to the property $sn^{-1}(-z;\kappa)=-sn^{-1}(z;\kappa)$. The profile for $\kappa< 1$ goes to zero as $x\rightarrow\pm\infty$;  however, the profile for $\kappa=1$ diverges as $x\rightarrow -\infty$. This profile must be compared with the profile at a maximum described in the figure (\ref{solatmin}).}  
  \label{solatminimun}
\end{center}
\end{figure}

The equation of motion in the form (\ref{soliton1}) can be solved for the field fluctuation $\Delta \phi$ around a minimum, considering the functional form (\ref{minex}); hence according to the Eq. (\ref{soliton1}),
\begin{eqnarray}
 \int \frac{ d(\Delta \phi)}{sn ( \Delta \phi; \kappa)} = \pm \sqrt{\frac{2}{1-c_{0}^2}} m(x - c_{0} t)\equiv \sigma\xi; 
\label{am1}
\end{eqnarray}
such an integral is known, namely,
$\int \frac{d\phi}{sn} =  \ln \frac{sn}{dn + cn}$; one can solve these expressions to favor of the field fluctuation $\Delta\phi$,
\begin{eqnarray}
 \Delta \phi (x, t ; \kappa) = sn^{-1} (\pm \frac{ 2 e^{\pm \sigma \xi} } {  \sqrt{ (\kappa - 1)^2 + 2 (\kappa + 1) e^{\pm2 \sigma \xi} + e^{\pm4\sigma \xi} }};\kappa)
\label{am3},
\end{eqnarray}
$sn^{-1}$ is the inverse of the elliptic function; note that this inverse function is described also by the same elliptic modulus that the original elliptic function $sn$. The signs in the argument are related by a mirror symmetry, namely, $sn^{-1}(-z;\kappa)=-sn^{-1}(z;\kappa)$. For a concrete description of this solution we choose the sign $+$ for both, the variable $\sigma\xi$, and for the argument in (\ref{am3}), and we assume that $m>0$. The choice $-$ for the variable $\sigma\xi$, can be constructed from the description with the choice $+$ by an interchanged 
of the roles of the coordinates $(x,t)$.
Under these considerations the profile of the solution for different values of the modulus are illustrated in the figure \ref{solatminimun}.

The cusp-like solitons have not been associated traditionally to the generalizations of Gordon field theory, instead they are found as solutions of a typical integrable system, namely, the Harry-Dym equation (see for example \cite{cusp} for more details). This atypical solution at the minima of our model, must be compared with the solitons found at the maxima for the second case below (see figure \ref{solatmin}), the so called kink-soliton typically associated with the Gordon field theory.

Although this cusp-like profile is maintained for all time, its height goes to zero as $t$ increases, due 
to the following time asymptotic limits
\begin{eqnarray}
\lim_{t \rightarrow \pm \infty}  \Delta \phi (x, t; \kappa) = 0;  \quad \kappa \neq 1;
\label {am4}
\end{eqnarray}
thus, the cusp-like traveling solution is stable.

For $\kappa=1$ the function $sn^{-1}$ reduces to the function $tanh^{-1}$, which leads to the following time asymptotic limits;
\begin{eqnarray}
\lim_{t \rightarrow  +\infty} \Delta \phi (x, t; \kappa=1) = 0; \quad \lim_{t \rightarrow  -\infty}   \Delta \phi (x, t; \kappa=1) = \infty;
\label{am44}
\end{eqnarray} 
The divergences for the field amplitude for $\kappa=1$, at $x\rightarrow-\infty$, and $t\rightarrow-\infty$, can be eliminated by choosing a different coordinate frame, namely, the null-coordinate frame, on which the field amplitude will be a constant for a given $\kappa$.

\subsubsection{null coordinate frame and the limit $\kappa\rightarrow 1^{-}$}
\label{null}
The null coordinate is defined as $\xi = 0=k_{0}(x - c_{0} t)$, for $k_{0}\neq 0$; in this frame the field amplitude (\ref{am3}) reduces to
\begin{eqnarray}
 \Delta \phi (\xi = 0; \kappa) = sn^{-1} (\pm \frac{ 2 } {  \sqrt{ k^2+4}};\kappa), \quad \kappa<1;
\label{null1}
\end{eqnarray}
hence, the field amplitude is constant in this frame; although
this expression is valid only in the weak regime ($\kappa<1$), it is finite
for the full range $\kappa\in R$; see the figure (\ref{cornull}) below. 
Furhtermore, note that the argument will have poles out of the real line, at $\kappa=\pm 2i$; however the function $sn^{-1}(\pm \infty;\kappa)$ is finite for $\kappa>1$ on the complex plane (see Eq. (\ref{finite}) below).
\begin{figure}[H]
  \begin{center}
   \includegraphics[width=.55\textwidth]{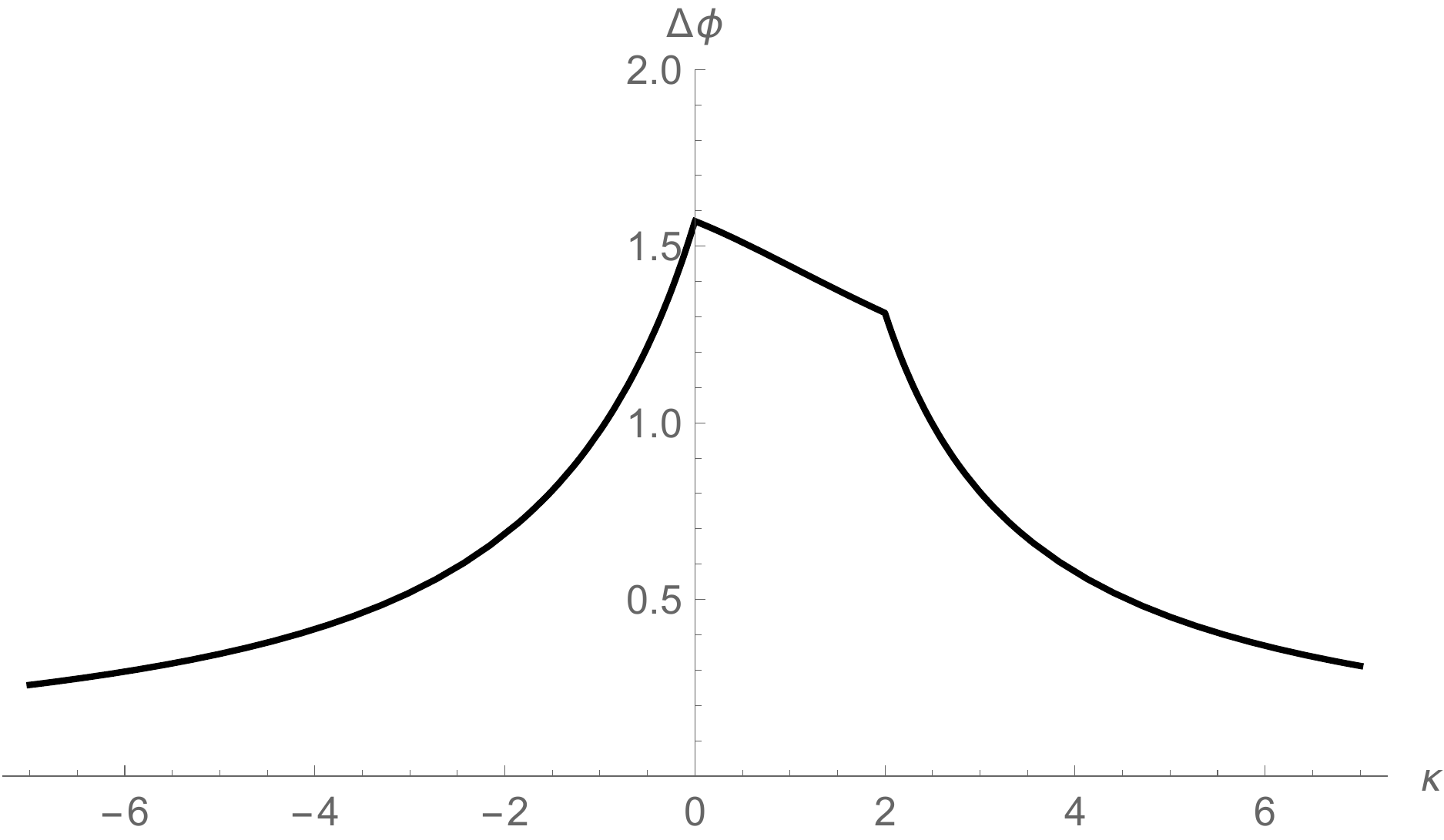}
  \caption{The field amplitude is finite for all $\kappa$, in particular for $\kappa=1$; note that the curve is not smooth for two self-dual points, $\kappa=0$, and  $\kappa=2$. Moreover, the field amplitude goes to zero as $\kappa\rightarrow \pm \infty$. } 
  \label{cornull}
\end{center}
\end{figure}
Let us see now that actually there exists an infinite number of Lorentz-frames with null coordinates; a $v$-Lorentz boost on the wave variable reads
\begin{eqnarray}
(x-c_{0}t)=\frac{1}{\sqrt{1-v^2}}\big[(1+vc_{0})x'-(v+c_{0})t'\big];
\label{lorentz}
\end{eqnarray}
hence the family of $v$-Lorentz frames with null coordinates are defined as,
\begin{eqnarray}
x'=(\frac{v+c_{0}}{1+c_{0}v})t';
\label{lorentz1}
\end{eqnarray}
these frames describe the field amplitude illustrated in the figure; for example with $v=0$ one recuperates the original frame. Moreover, the expression (\ref{lorentz}) will allow us to find different Lorentz frames from which the divergences around the self-dual point $\kappa=1$ can be controlled (see section \ref{nonull1}).

\subsubsection{A complex solution for $\kappa\in R$}
\label{complexsol1}
\begin{figure}[H]
  \begin{center}
   \includegraphics[width=.55\textwidth]{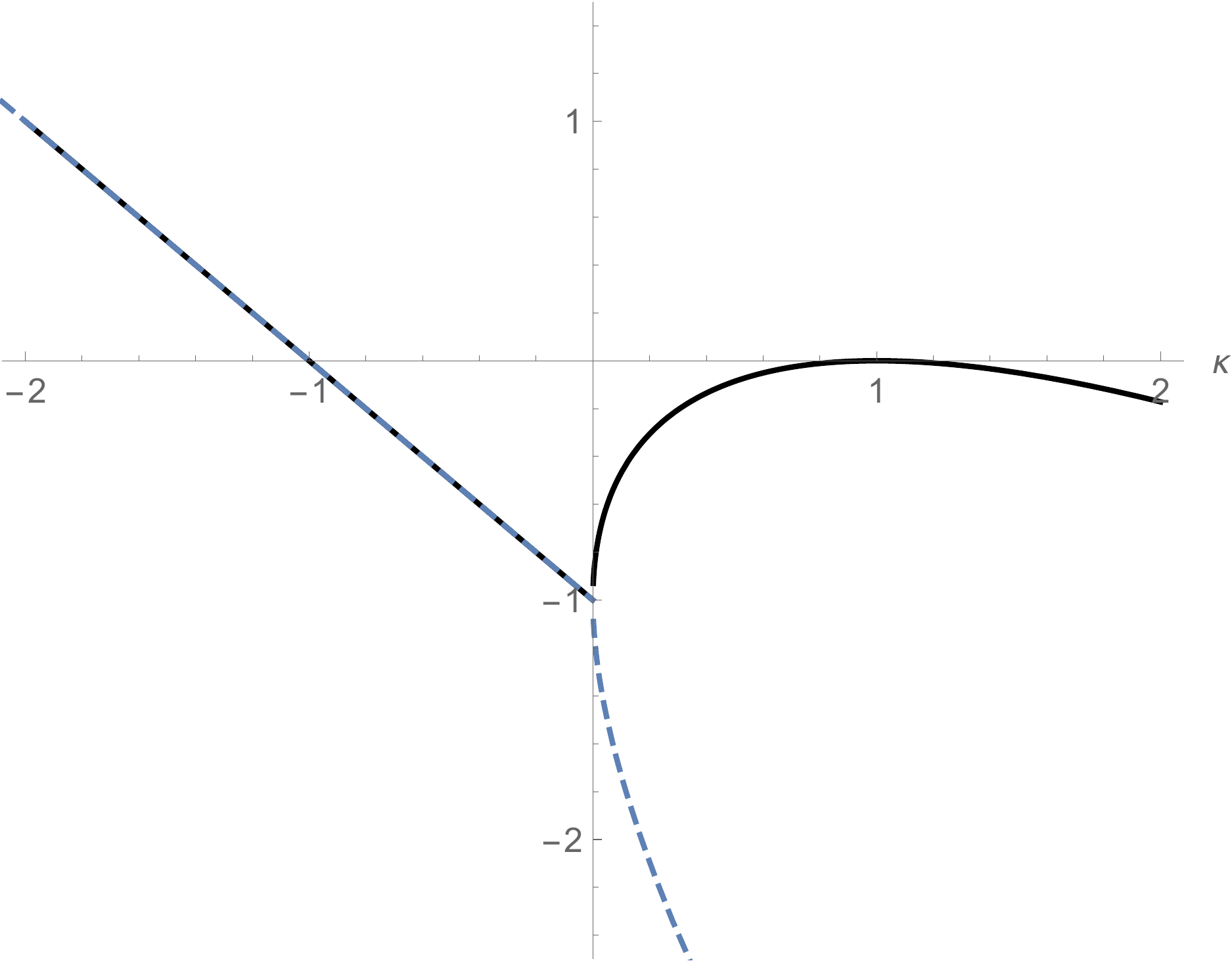}
  \caption{The real parts of the expressions $-(\sqrt{k}\pm 1)^2$, for $\kappa\in R$; the
   black and dashed curves correspond to the choice $-$, and $+$ respectively; the curves coincide to each other along the straight line.}  
  \label{singu1}
\end{center}
\end{figure}

\begin{figure}[H]
  \begin{center}
   \includegraphics[width=.55\textwidth]{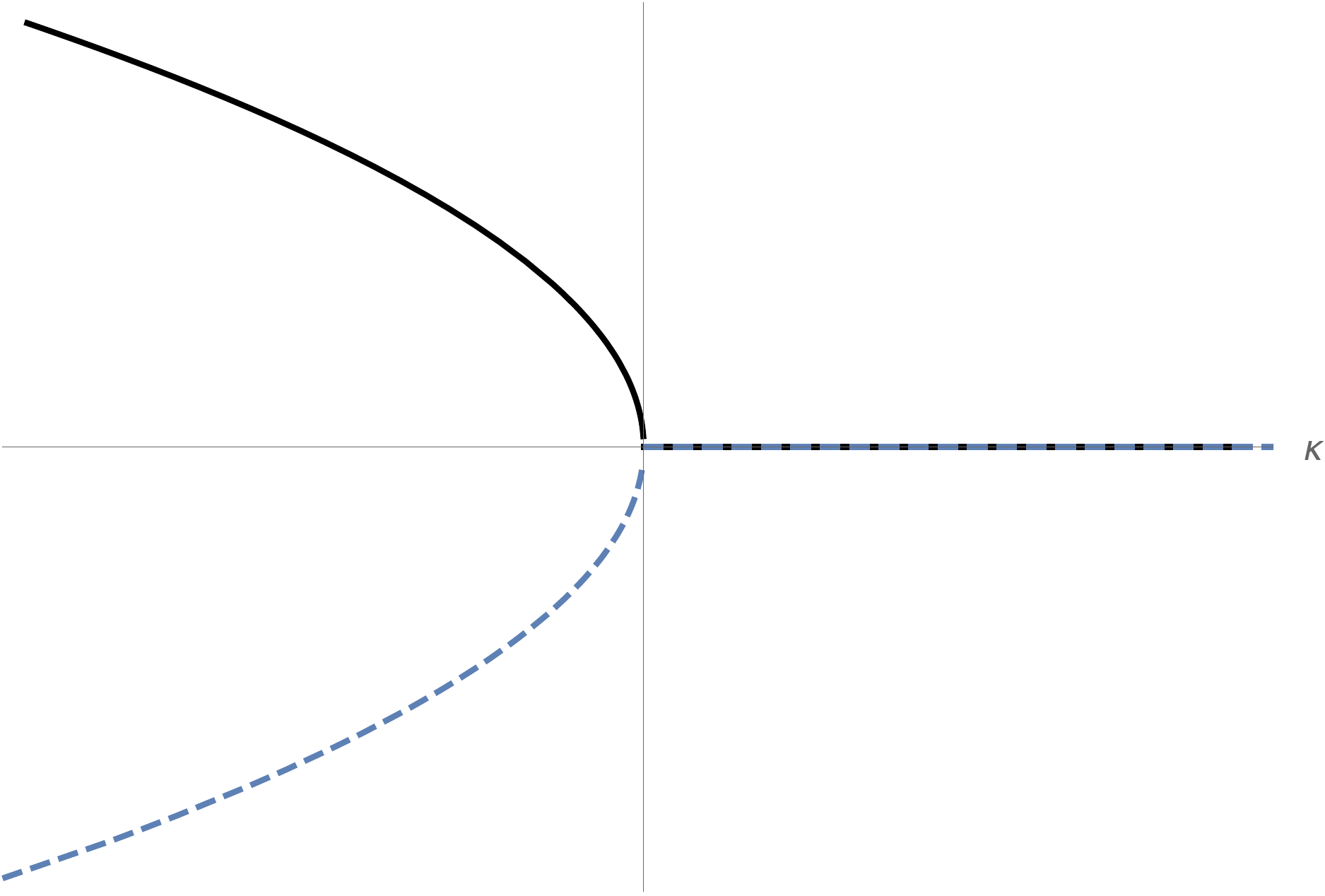}
  \caption{The imaginary parts of the expressions $-(\sqrt{k}\pm 1)^2$; the
   black and dashed curves correspond to the choice $-$, and $+$ respectively; the curves coincide to each other along the straight line with $\kappa\geq 0$.}  
  \label{singu2}
\end{center}
\end{figure}

The argument for field amplitude (\ref{am3}) diverges provided that
\begin{eqnarray}
	e^ {2 \sigma \xi} = - (\sqrt{\kappa} \pm 1)^2;
\label{diver1}
\end{eqnarray}
in particular for $\kappa=1$ and for the negative branch one obtains the null coordinate frame scenario des\-cribed previously. Thus, for $\kappa\neq 1$ we have the complex exponentials as the only solutions; hence, whereas the exponential $e^ {2 \sigma \xi}$ is real, the argument in the Eq. (\ref{am3}) does not diverge for $\kappa \in R-\{1\}$. The real and imaginary parts of the roots are illustrated in the figures (\ref{singu1}) and (\ref{singu2}) for $\kappa\in R$. Considering the definition (\ref{am1}) for $\sigma\xi$, the complexification of $e^ {2 \sigma \xi}$ implies that, {\it i)}: the mass is imaginary $m\rightarrow im$; {\it ii)}: $c_{0}>1$, which implies that the soliton propagates faster than the light velocity; {\it iii)}: the space-time coordinate $x-c_{0}t$ is complex; this fact will imply a space-time oscillatory  behavior at the minima. We consider that the later case is the only physically meaningful in the present discussion.

Although the argument diverges according to the expression (\ref{diver1}), the field amplitude itself is finite through an analytical continuation provided that $\kappa>1$,
\begin{eqnarray}
 \Delta \phi (\pm \infty; \kappa) = sn^{-1} (\pm \infty;\kappa)=\pm [K(k)-\frac{1}{\sqrt{k}}K(\frac{1}{k})]=\pm i K(i\sqrt{\kappa^2-1}),\quad \kappa>1.
\label{finite}
\end{eqnarray}

\subsection{The solution general for the equation of motion}
\label{sntanh}
The general solution for the equation (\ref{soliton1}) for the potential $V_{1}$ can be constructed  by using the expression (\ref{ei2}), along the lines that we followed for finding the solution (\ref{am3});
the field amplitude will read
\begin{eqnarray}
\phi(x,t;\kappa) & = & sn^{-1}\big[ tanh(\pm\sigma\xi);\kappa  \big]\label{gensol1}\\
& = & z+\frac{1}{6}(\kappa+1)z^3+\frac{1}{40}(3\kappa^2+2\kappa+3)z^5+....;\quad z\equiv  tanh(\pm\sigma\xi);
\label{exptanh1}
\end{eqnarray}
where we have displayed the Taylor expansion, which corresponds to an expansion in terms of the $tanh$-function;
in particular the first order term gives the usual solution for a soliton described by the $tanh$ -function; considering that $sn^{-1}(\theta,\kappa=1)=tanh^{-1}\theta$, then the solution is trivial at the self-dual point, since $\phi(x,t;\kappa=1) =\pm\sigma\xi$; this case separates nontrivial cases with $\kappa<1$ and for $\kappa>1$. Furthermore, 
 the Taylor expansion shows a critical point at $\kappa=-1$, since the terms of orders $3,7,11,$,etc, in $z$, vanish at such a self-dual point; thus, in this case only the terms of orders 1,5,9, etc, will be present.
 
 The tanh-functions profile is identified with the so called kink solitons, whose spatial configuration is illustrated in the figure \ref{solatmin};  typically these soliton configurations take the vev at two vacua as boundary conditions.
 For the weak coupling regime ($\kappa<1$), with finite barriers between vacua for the full potential illustrated in the figure \ref{cn}, we can consider the following boundary conditions 
\begin{eqnarray}
\phi(-\infty,\kappa)=2n_{-}K(\kappa), \quad \phi(+\infty,\kappa)=2n_{+}K(\kappa);
\label{bc}
\end{eqnarray}
due to the periodicity is defined by $2K(\kappa)$; thus the winding number is given as usual by $Q=n_{+}-n_{-}$, and the corresponding soliton is characterized by this topological number.

\subsubsection{Expansion around the maxima: a kink-like soliton will decay to a cusp-like soliton}
\label{kinkcusp}

In analogy with the expansion around minima, the expansion around maxima will read
\begin{eqnarray}
V_{1}(max+\Delta\phi;\kappa)=V_{1}(\Delta\phi;\kappa),\quad k<1;
\label{aroundmax}
\end{eqnarray}
it means there is not a functional change in the potential respect to the original expression; this result can be proved by using the addition formulae and the fact that  
 the period is $2K(\kappa)$, which coincide with the location of the maxima in the potential (see Eq. (\ref{maxmin})). Hence, the expression (\ref{gensol1}) can be considered as the solution for the field fluctuation $\Delta\phi$ around the maxima; the general profile for this soliton looks like a kink solution (described in the figure (\ref{solatmin}) below) in the interval $\kappa<1$. 
Furthermore, the approximation of the expression (\ref{gensol1}) in the form $sn(\phi;\kappa)\approx \Delta\phi-\frac{1}{6}(\kappa+1)(\Delta\phi)^3=tanh(\pm\sigma\xi)$, valid in the expansion around the maxima, will lead also to kink-like solution, for both, for the linear and cubic approach for the fluctuation $\Delta\phi$.
Therefore, if this kink-like solution (at the maxima) will decay to its vacuum configuration (at the minima) described previously, then will decay to a cusp-like configuration.

For this high-amplitude field decay beyond the quartic field theory approach around the maxima and the minima, one can compare directly the corresponding mass quadratic terms in the maximum at the point $\phi=0$, and in the first minimum 
at $\phi=K(\kappa)$, for example. Note first that the mass quadratic term in the Eq. (\ref{minex}) is independent on the modulus $\kappa$ (with the correct sign); however, the mass term in the expression (\ref{expan1}) is depending on $\kappa$, in fact it is critical at the self-dual point with $\kappa\rightarrow 1^{-}$, for which the particle is light. This fact does not affect the mass generated by SSB at the minimum, contrary to the mass generated in the quartic field theory at the small-amplitude approach, according to the Eq. (\ref{genmass}).

\subsection{$\kappa>1$}
\label{kappabig}

\begin{figure}[H]
  \begin{center}
   \includegraphics[width=.55\textwidth]{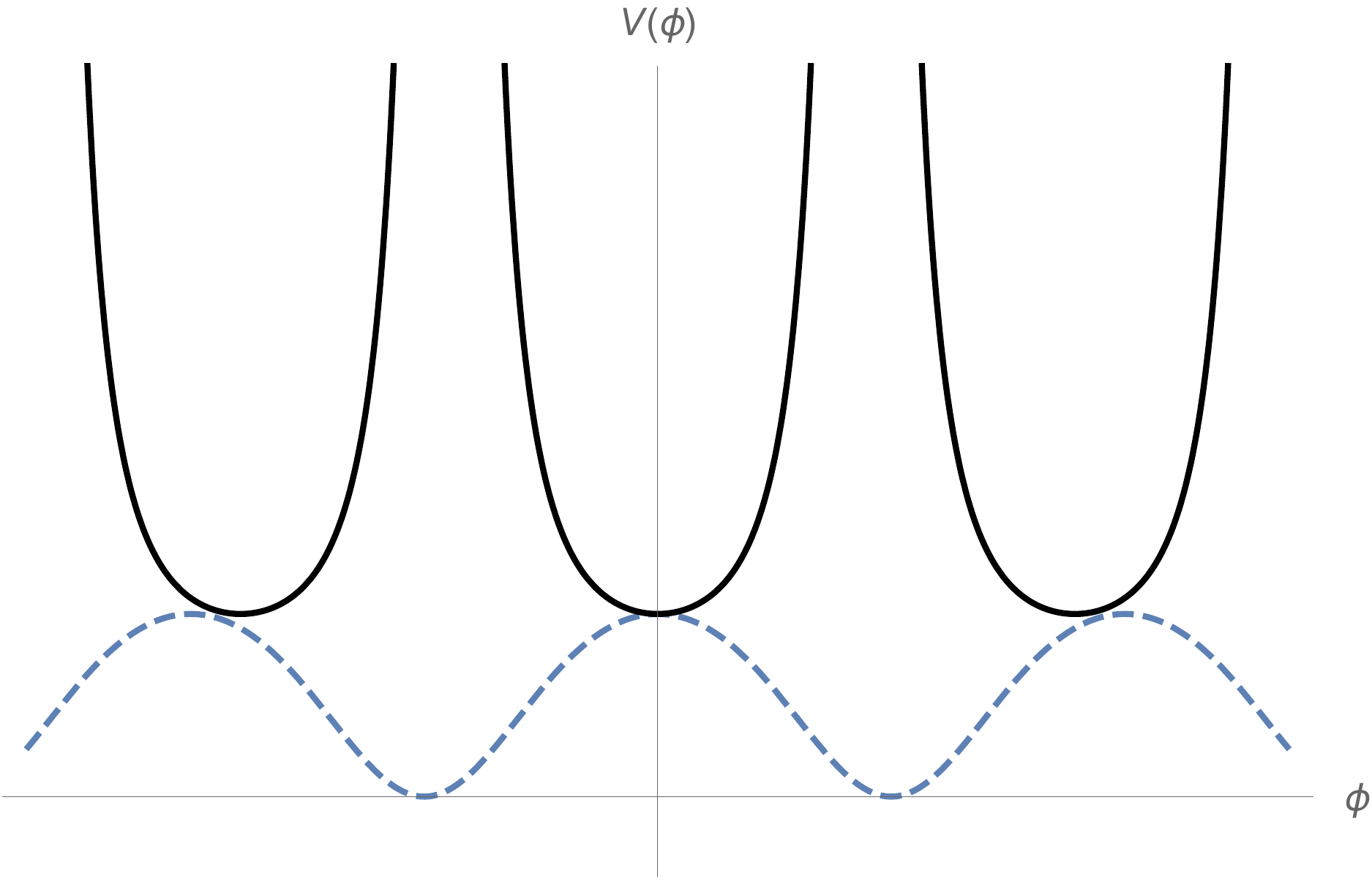}
  \caption{The potential (\ref{potlimit}) for $\kappa>1$ corresponds to an infinite collection of disconnected vacua, without possibility of tunneling between them; the dashed curve represents the same potential for $\kappa<1$, and described in the figure \ref{cn} .  Hence, there exists a change in the topology of the potential around $\kappa\rightarrow1$.  According to the Eq. (\ref{potlimit}) the potential is constant just for $\kappa=1$.}  
  \label{mmcn}
\end{center}
\end{figure}

From the beginning the self-dual point $\kappa=1$ has shown to be special; first, the potential $V_{1}$ is trivial for $\kappa=1$ according to the Eq. (\ref{potlimit}); however,
the signs of the derivatives in the expressions (\ref{der}), depend sensitively on the limits from the left and from the right, $\kappa\rightarrow 1^\pm$. Specifically the second derivative expanded around  $\kappa=1$ will read
\begin{eqnarray}
\frac{d^2V_{1}}{d^2 \phi}\approx 2m^2 (\kappa-1)\big(1+2\sinh^2(\phi) \big)+\mathcal{O}(\kappa-1);
\label{sde}
\end{eqnarray}
thus $\frac{d^2V_{1}}{d^2 \phi}>0$ as $\kappa\rightarrow 1^+$, and $\frac{d^2V_{1}}{d^2 \phi}<0$ as $\kappa\rightarrow 1^-$, for arbitrary $\phi$. For example for $\phi=0$ we have a minimum as  $\kappa\rightarrow 1^+$, and a (relative) maximum as  $\kappa\rightarrow 1^-$ (see the figure \ref{mmcn}).

The change of topology in the potentials described in the figure implies a change of topology in the vacuum manifold; for the strong coupling regime ($\kappa>1$) the vacua correspond to an infinite collection of single points, with infinite barriers between them; thus the vacuum manifold will have not homotopy constraints, and there is not a possibility of formation of topological defects.  For the weak coupling regime ($\kappa<1$), with finite barriers between vacua, the boundary conditions (\ref{bc}) are imposed on the general solution (\ref{gensol1}).

\section{Second case}
\label{sc}
The second case consists of the identification of the integrant $m/ \sqrt{V(\phi)}$  in the Eq. (\ref{soliton1}), with the integrant in the Eq. (\ref{ei1}); the potential takes the form 
\begin{eqnarray}
V_{2}(\phi, \kappa)=m^2\frac{sn^2(\phi, \kappa)}{dn^2(\phi, \kappa)},  \quad
V_{2}(\phi, \kappa=0)= \frac{m^2}{2}(1-\cos 2\phi); \quad
V_{2}(\phi, \kappa=1)=\frac{m^2}{2}\sinh^2 \phi;
\label{pot2}
\end{eqnarray}
similarly the expansion around $\phi =0$ reads
\begin{eqnarray}
V_{2}(\phi, \kappa)=m^2+m^2\phi^2 + \frac{2}{3}m^2(\kappa-\frac{1}{2})\phi^4+\mathcal{O}(\phi^6)+...
\label{expand2}
\end{eqnarray}
however, as opposed to the previous case, the point $\phi =0$ is a minimum (see the figure \ref{mmsn}); in fact the mass term in the above expression has already the correct sign such as that generated by SSB mechanism, and it is independent on the elliptic modulus. Moreover, in similarity with the previous case, the quartic term is vanishing at $\kappa\rightarrow 1/2$, and the coefficient for the sixth order term 
will have the form $C_{6}(\kappa)\equiv \frac{1}{45}[17\kappa(\kappa-1)+2]$, with $C_{6}(\kappa=1/2)<0$, obtaining an "up side down"
SSB scenario. In this manner, one has to expand the potential $V_{2}$ around a maximum, in order to obtain a standard (polynomial) SSB scenario.

Along the same lines of the previous case, the role of maxima and minima is interchanged respect to the description given in Eq.
(\ref{maxmin}),
\begin{eqnarray}
      \phi_{critical} = \left\{
    \begin{array}{cc}
      <\phi>  \equiv 2nK(\kappa),  & \textrm{Minima};\\   \phi_{_{M}}\equiv(2n+1)K(\kappa), & \textrm{Maxima};
    \end{array}
\right.  \label{minmax}
\end{eqnarray}
again, we are considering that $\kappa<1$, since the profile for $V_{2}$ with its maxima and minima is possible only with this restriction, due to the limit  $lim_{\kappa\rightarrow 1^-} K(\kappa)=\infty$.

The corresponding series expansions around arbitrary maxima and minima are given by,
\begin{eqnarray}
    V_{2}(\phi,\kappa)|_{_{max}} = -\frac{m^2}{  \kappa -1}+\frac{m^2 }{\kappa -1}\left(\phi- \phi_{_{M}}\right)^2-\frac{  (\kappa +1) m^2 }{3 (\kappa -1)}\left(\phi- \phi_{_{M}}\right)^4+\ldots;\label{expan4}
\end{eqnarray}
\begin{eqnarray}
     V_{2}(\phi,\kappa)|_{_{min}} =
     m^2 \left(\phi-<\phi>\right)^2+\frac{2}{3}   (\kappa -\frac{1}{2}) m^2 \left(\phi-<\phi>\right)^4+ \ldots;\label{expan5}
\end{eqnarray}
For the expansion (\ref{expan5})
 the coefficients for the terms of orders $4,8,12,...,$ in $\phi-<\phi>$, vanish at an unique self-dual point, 
$\kappa=1/2$. Moreover, for the expansion (\ref{expan4}) all the coefficients diverge at the self-dual point $\kappa=1$; however the potential itself is well defined for $\kappa=1$ (Eq. (\ref{pot2})). Additionally for the expansion  (\ref{expan4}) the coefficients for the terms of orders $4,8,12,...,$ in $\phi-\phi_{M}$, vanish for an unique self-dual point, 
$\kappa=-1$.

\subsection{Expanding around a maximum: A kink-like soliton}
\label{kinksoliton}
A functional form for the potential can be constructed around a maximum by using again the addition formulae for the elliptic functions
\begin{eqnarray}
V_{2}(max+\Delta\phi;\kappa)=m^2\frac{cn^2(\Delta\phi;\kappa)}{1-\kappa}\approx\frac{m^2}{(1-\kappa)}(1-\Delta\phi^2+\frac{\kappa+1}{3}\Delta\phi^4+\ldots);
\label{maxex}
\end{eqnarray}
the Taylor expansion coincides with the general expansion (\ref{expan4}) with $\Delta\phi=\phi-\phi_{M}$.

Similarly the integration of the equation of motion for the potential (\ref{maxex}), by using the expression $\int \frac{dz}{cnz} = \frac{1}{\sqrt{1-\kappa}} ln \frac{dnz + \sqrt{1 -\kappa}snz}{cn z}$, leads to
\begin{eqnarray}
 \Delta \phi (x, t ; \kappa) = sn^{-1} [\ \pm \frac{e^{2 \sigma \xi} - 1}{\sqrt {(e^{2\sigma \xi}+1)^2 -4 \kappa e^{2\sigma \xi}}} ;\kappa]\ ;
\label {am5}
\end{eqnarray} 
for large $x$, and $t$ the field tends to a constant value defined by the complete elliptic integral,
\begin{eqnarray}
\lim_{t \rightarrow \pm \infty}  \Delta \phi = \mp K(\kappa),\quad \lim_{x \rightarrow \pm \infty}  \Delta \phi = \pm K(\kappa);\quad
\lim _{\kappa \rightarrow 1}K(\kappa)=\infty;
\label {am6}
\end{eqnarray} 
where we have considered that $\sigma > 0$, which means $m > 0$; moreover for $\sigma < 0$ the signs in the above expression are interchanged. 

\begin{figure}[H]
  \begin{center}
   \includegraphics[width=.55\textwidth]{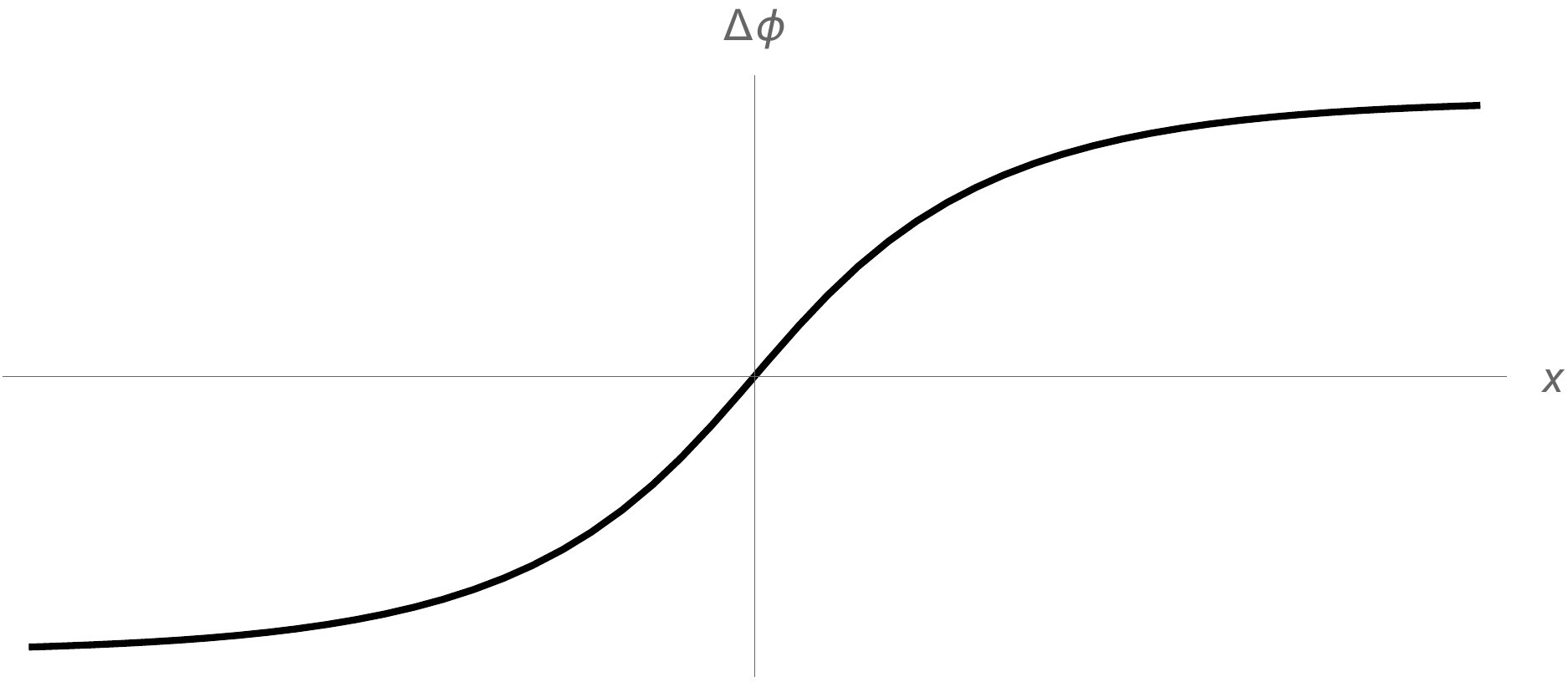}
  \caption{The kink-like profile as the spatial configuration of the soliton for $t=0$ at a maximum, for $\kappa<1$; note the asymptotic limits as $x\rightarrow\pm\infty$ according to the Eq. (\ref{am6}).}  
  \label{solatmin}
\end{center}
\end{figure}

\begin{figure}[H]
  \begin{center}
   \includegraphics[width=.55\textwidth]{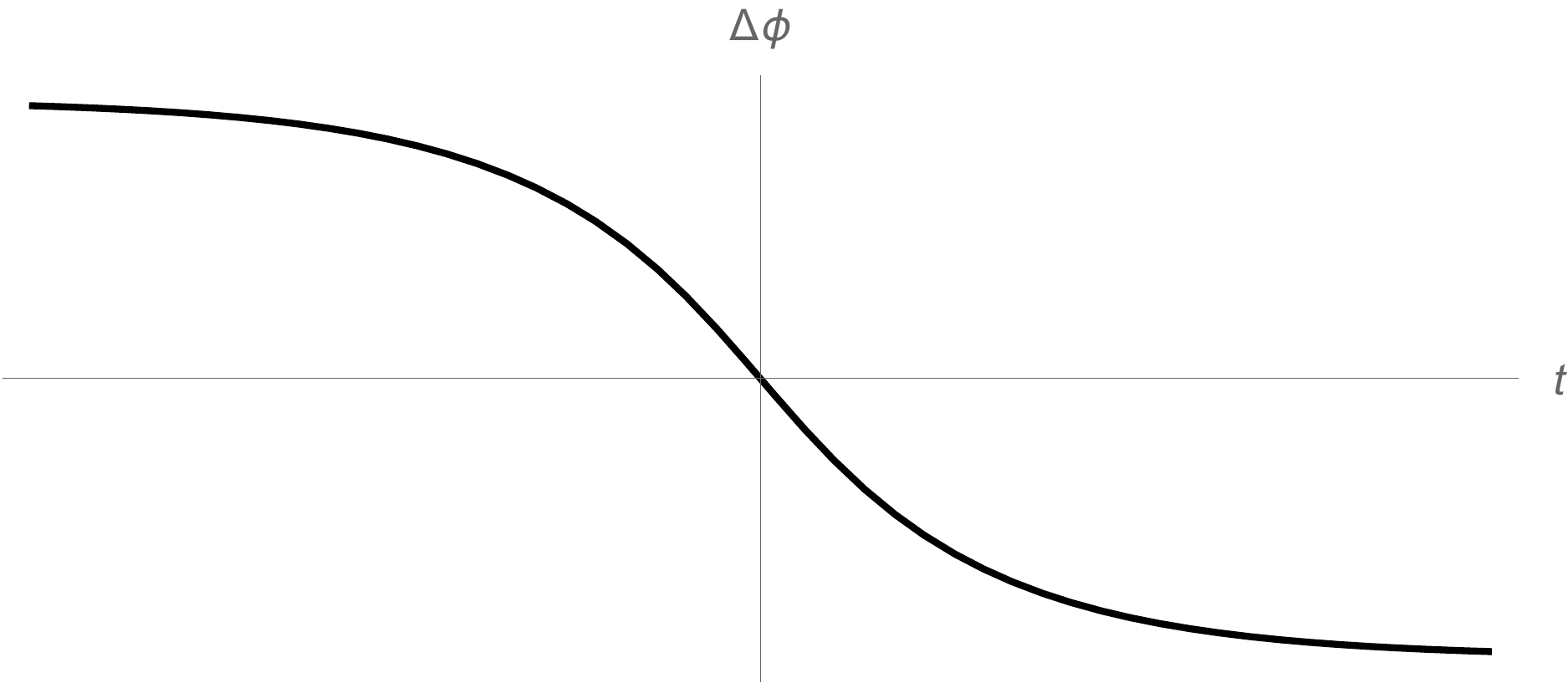}
  \caption{The (real) field amplitude as a function on $t$ for arbitrary fixed $x$-coordinate; for  $ \kappa <1$ there no exist singularities in the evolution; note the asymptotic limits (\ref{am6}).}  
  \label{solatmin1}
\end{center}
\end{figure}

\subsubsection{the null coordinate frame and the limit $\kappa\rightarrow 1$}
\label{nonull1}
\begin{figure}[H]
  \begin{center}
   \includegraphics[width=.55\textwidth]{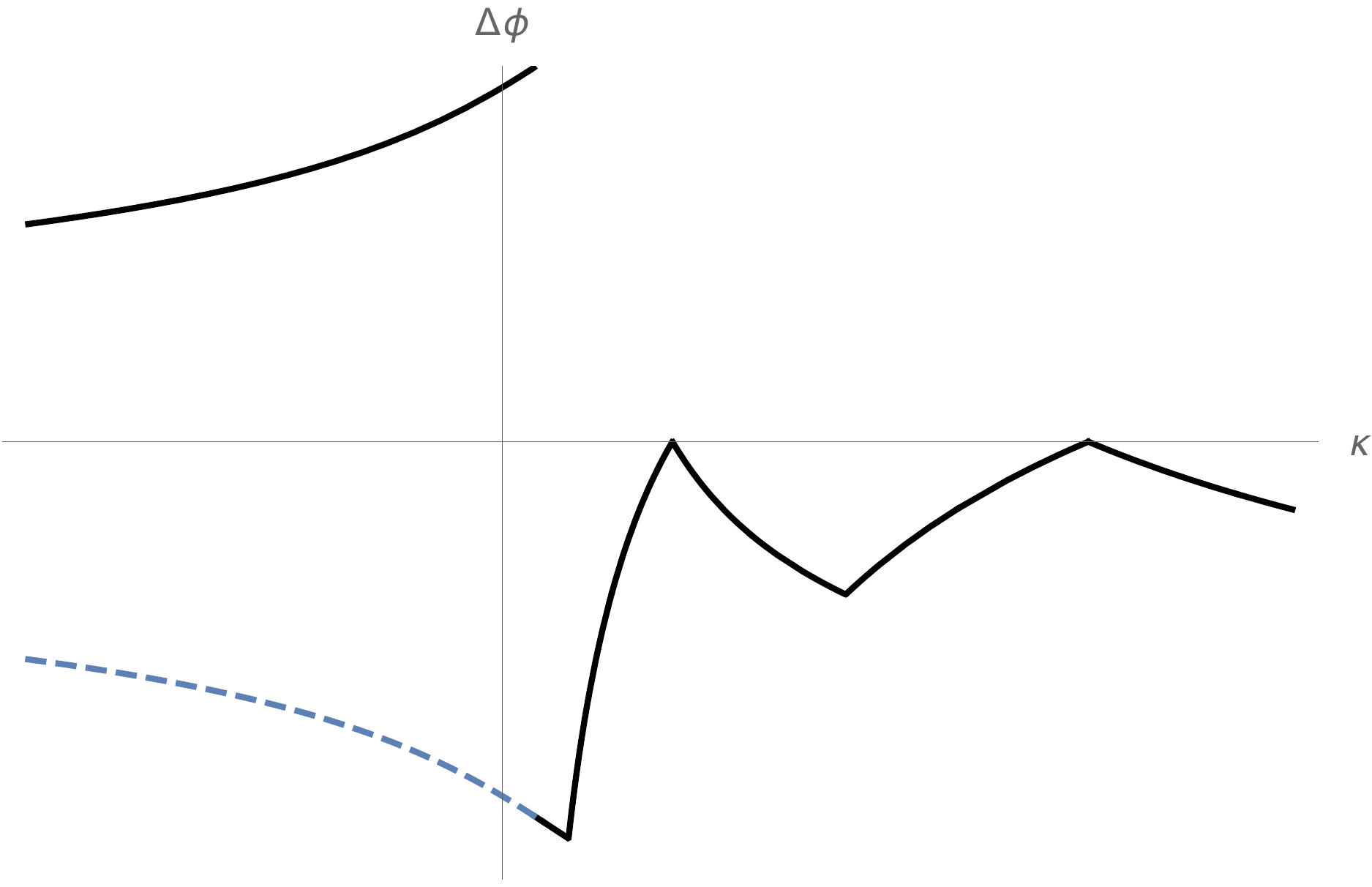}
  \caption{The field amplitude is finite for all $\kappa$. The stepped black curve represents the case with $a>0$;  the limits (\ref{nnc5}) are shown; note that it is smooth for $\kappa<1$, and it is a smooth piecewise curve for $\kappa>1$. For $a<0$ the field is shown as a dashed curve for $\kappa<1$, which corresponds to the mirror image on the $\kappa$-axis of the black curve for $\kappa<1$; for $\kappa>1$ the dashed curve coincides with the black curve in this interval, and thus one has the continuous limit (\ref{nnc6}); thus, the complete dashed (continuous) curve is the wanted finite description for the field amplitude in the $\kappa$-parameter space. Moreover, the field amplitude goes to zero as $\kappa\rightarrow \pm \infty$. } 
  \label{piecewise}
\end{center}
\end{figure}

We consider by simplicity the $+$ sign in the argument in the solution (\ref{am5}); the another case can be obtained by the property
$sn^{-1}(-z;\kappa)=-sn^{-1}(z;\kappa)$.
The asymptotic limits (\ref{am6}) diverge as lim $K(\kappa\rightarrow 1)=\infty$; as it was shown in the section \ref{null}, one may use the null coordinate for controlling such divergences. For the case at hand this trick does not work; first we note that 
\begin{eqnarray}
\lim_{\xi \rightarrow 0} sn^{-1} [\ \pm \frac{e^{2 \sigma \xi} - 1}{\sqrt {(e^{2\sigma \xi}+1)^2 -4 \kappa e^{2\sigma \xi}}} ;\kappa]=0, \quad {\rm for} \quad \kappa\neq 1;
\label{nnc1}
\end{eqnarray}
in fact, around $\xi=0$, we have,
\begin{eqnarray}
\frac{e^{2 \sigma \xi} - 1}{\sqrt {(e^{2\sigma \xi}+1)^2 -4 \kappa e^{2\sigma \xi}}}\approx\frac{  \sigma\xi}{\sqrt{1-\kappa}};
\quad 
sn^{-1} [\  \frac{e^{2 \sigma \xi} - 1}{\sqrt {(e^{2\sigma \xi}+1)^2 -4 \kappa e^{2\sigma \xi}}} ;\kappa]\approx\frac{  \sigma\xi}{\sqrt{1-\kappa}};
\label{nnc2}
\end{eqnarray}
the coincidence in these first order approximations  is due to the expansion $sn^{-1}(z;k)\approx z$; 
the above expressions diverge at $\kappa\rightarrow 1$; note that the limits $\kappa\rightarrow 1^{+}$, and $\kappa\rightarrow 1^{-}$ are qualitatively different. Moreover, by fixing $\kappa=1$ from the beginning in the solution (\ref{am5}), then
\begin{eqnarray}
\lim_{\xi \rightarrow 0^{\pm}} tanh^{-1} [ \frac{e^{2 \sigma \xi} - 1}{\sqrt {(e^{2\sigma \xi}+1)^2 -4  e^{2\sigma \xi}}}]=\pm\infty, \quad {\rm for} \quad \kappa=1.
\label{nnc3}
\end{eqnarray}
However, this infinite-step field configuration around $\kappa=1$ can be regularized by choosing  a different coordinate frame; the expression (\ref{nnc2})  suggests a different coordinate frame,
\begin{eqnarray}
 2\sigma\xi=a\sqrt{1-\kappa};
 \label{nnc4}
\end{eqnarray}
from this frame we can construct, depending on the sign of the proportionality constant $a$, two finite solutions, namely, 
a finite stepped function, and a continuous function on the all of range for the parameter $\kappa$. For the former we have the following behavior at $\kappa=1$,
\begin{eqnarray}
\lim_{\kappa \rightarrow 1^{\pm}} sn^{-1} [\  \frac{e^{a\sqrt{1-\kappa}} - 1}{\sqrt {(e^{a\sqrt{1-\kappa}}+1)^2 -4 \kappa e^{a\sqrt{1-\kappa}}}} ;\kappa]=\mp tanh^{-1}\big(\frac{a}{\sqrt{a^2+4}}\big), \quad {\rm for} \quad a>0;
\label{nnc5}
\end{eqnarray}
and for the later we have the same limit from both sides,
\begin{eqnarray}
\lim_{\kappa \rightarrow 1} sn^{-1} [\  \frac{e^{a\sqrt{1-\kappa}} - 1}{\sqrt {(e^{a\sqrt{1-\kappa}}+1)^2 -4 \kappa e^{a\sqrt{1-\kappa}}}} ;\kappa]=tanh^{-1}\big(\frac{a}{\sqrt{a^2+4}}\big), \quad {\rm for} \quad a<0;
\label{nnc6}
\end{eqnarray}
both solutions are shown in the figure \ref{piecewise}.
According to the expression (\ref{lorentz}), from a general Lorentz frame, the identification (\ref{nnc4}) will read
\begin{eqnarray}
\frac{2\sigma}{\sqrt{1-{v^2}}}\big[(1+{vc_{0}})x'-(v+c_{0})t'\big]=a\sqrt{1-\kappa};
\label{lorentz4}
\end{eqnarray}

Of course there exists an infinite number of frames, and we have described a way to find out an appropriate frame for controlling the divergences in the solutions. In fact the left-hand side in the above equation can be identified in general with an arbitrary function, $f(\kappa)$; for example in the first case described in Eq. (\ref{lorentz1}), $f(\kappa)=0$; one always can to find an appropriate $f$ to regularize the divergences of the solutions and, according to our results, a finite description for the solution can be obtained. However,  the resulting configuration, although finite and continuous, it is not smooth in a finite number of points (figure (\ref{cornull}) for the first case) or infinite number of points (figure (\ref{piecewise}) for the second case).

\subsubsection{A complex solution for $\kappa\in R$}
\label{complexsol2}
The argument in Eq. (\ref{am5}) diverges at 
\begin{eqnarray}
e^{2 \sigma \xi} = 2 \kappa - 1 \pm 2 \sqrt{ \kappa (\kappa -1)};
\label {singular}
\end{eqnarray} 
this equation has not real solutions for $e^{2 \sigma \xi}$ in the interval $\kappa < 1$, 
thus, the solution (\ref{am5}) is well behaved just in the interval where $V_{2}$ has the profile with its maxima and minima.
Moreover,  the above expression has real solutions in the interval $\kappa \geq 1$ (see figure (\ref{pole}) below); however for each $\kappa > 1$ the field amplitude is complex and finite, according to the property (\ref{finite}). For $\kappa=1$ the expression (\ref{singular}) leads, as a solution, to the null coordinate, $\xi=0$ (see section \ref{null}).

\begin{figure}[H]
  \begin{center}
   \includegraphics[width=.55\textwidth]{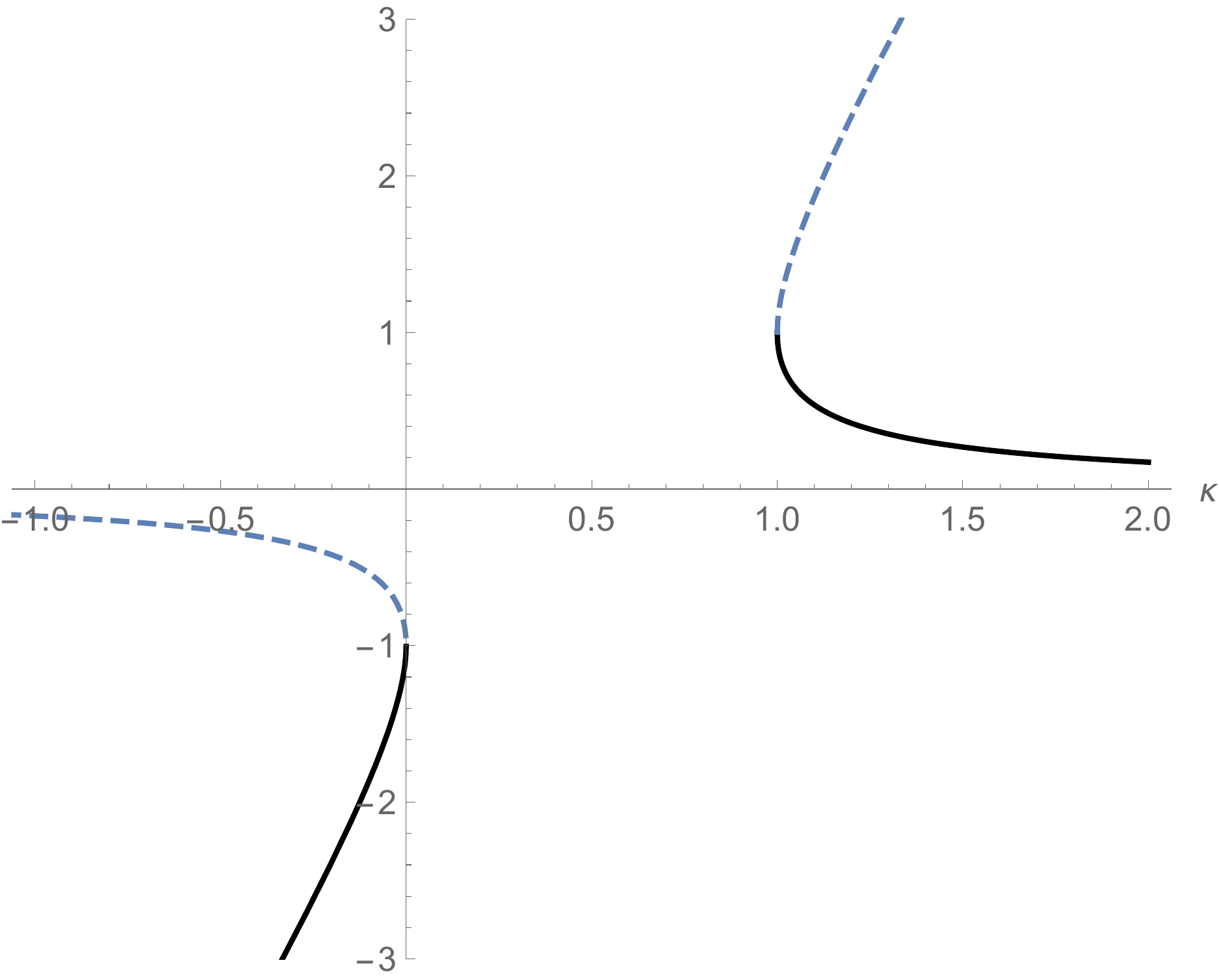}
  \caption{The black (disconnected) curves represent the expression $2\kappa-1-2\sqrt{\kappa(\kappa-1)}$ , and the dashed curves represent the expression  $2\kappa-1+2\sqrt{\kappa(\kappa-1)}$; for $\kappa\geq1$ there exist two branches in the first quadrant with positive values, for which the expression (\ref{singular}) have real roots for $\sigma\xi$; note that the branches form a continuous curve. Note also that in the interval $0 \leq \kappa <1$ there no exist real roots for $\sigma\xi$, and the solution (\ref{am5}) is not singular for real $\kappa $ within such an interval.}  
  \label{pole}
\end{center}
\end{figure}
\begin{figure}[H]
  \begin{center}
   \includegraphics[width=.55\textwidth]{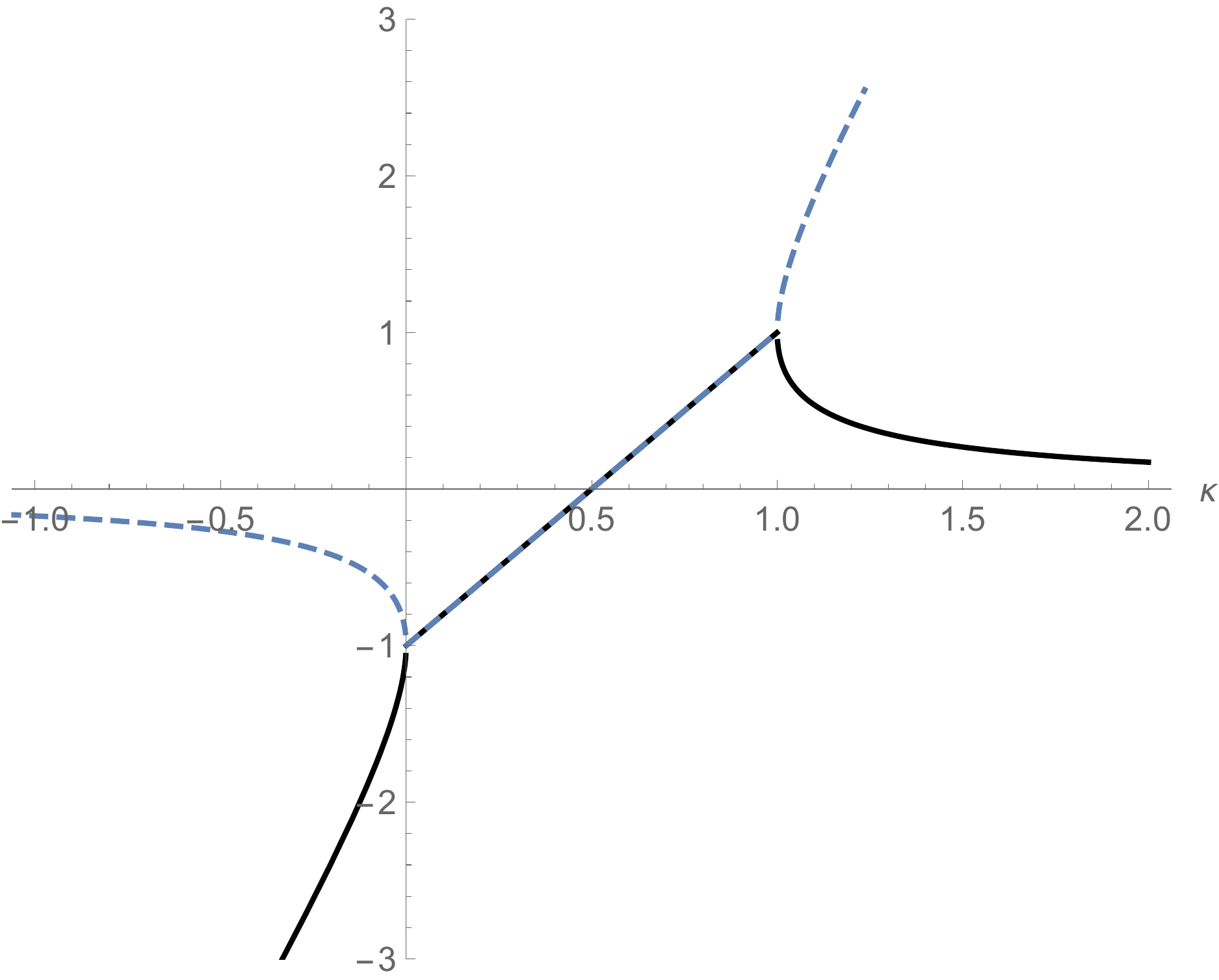}
  \caption{The black (now connected) curve represents the real part of the expression $2\kappa-1-2\sqrt{\kappa(\kappa-1)}$, including values in the interval $0\leq \kappa \leq 1$; similarly
 the dashed (connected) curve represents the real part of the expression  $2\kappa-1+2\sqrt{\kappa(\kappa-1)}$. These connected curves coincide to each other along the straight line 
 crossing the self-dual point $(1/2,0)$, and connecting the matching points.}  
  \label{loop1}
\end{center}
\end{figure}
\begin{figure}[H]
  \begin{center}
   \includegraphics[width=.55\textwidth]{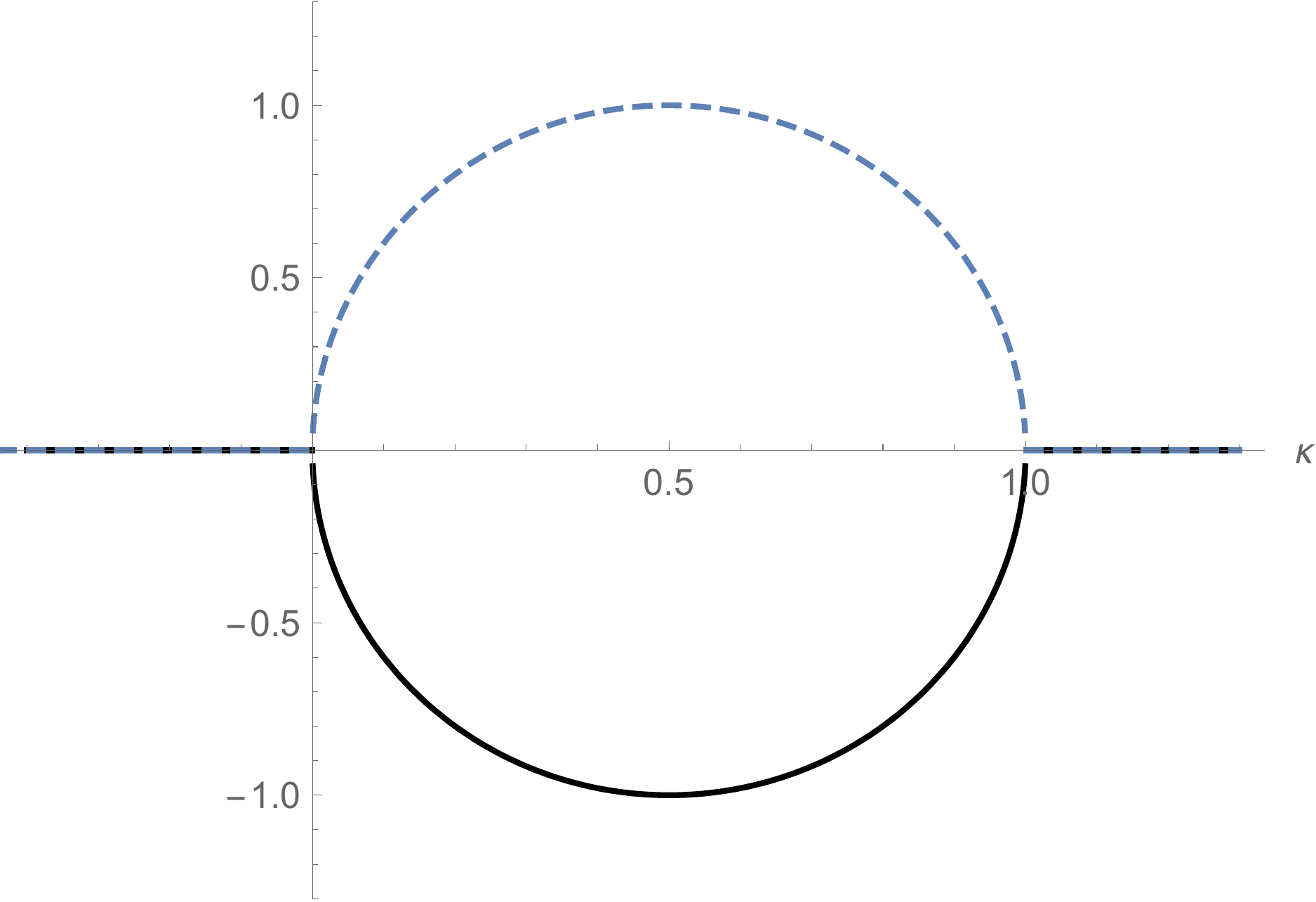}
  \caption{The black curve represents the imaginary part of the expression $2\kappa-1-2\sqrt{\kappa(\kappa-1)}$, and the dashed curve represents the imaginary part of the expression  $2\kappa-1+2\sqrt{\kappa(\kappa-1)}$, including values in the interval $0\leq \kappa \leq 1$. The curves coincide to each other along the $\kappa$-axis;
 note that the nontrivial branches form now a continuous loop.}  
  \label{loop2}
\end{center}
\end{figure}

If we admit complex roots for the singularities described in (\ref{singular}), then the values for $\kappa<1$ can be incorporated, in particular for $0 \leqslant \kappa < 1$.
The real and imaginary part of these roots are shown in the figures (\ref{loop1}) and (\ref{loop2}).
Therefore, for each value of $\kappa$ in the reals, there exist in general two complex roots for the Eq. (\ref{singular}), which reduce to two real solutions for $\kappa > 1$.
We recall that by admitting complex roots, we are admitting the complexification of the solution, and thus the scenarios described previously for the first case are possible (see section \ref{complexsol1}).


Note that the interval $0<\kappa<1$ originally prohibited in the figure \ref{pole}, now is admissible with the complexification on the complex plane; at the end the real and imaginary parts are now connected just on such an interval.

\subsection{The solution general for the equation of motion}
The general solution for the equation (\ref{soliton1}) for the potential $V_{2}$ can be constructed  by using the expression (\ref{ei1}), along the lines that we followed for finding the solution (\ref{gensol1});
the field amplitude will read
\begin{eqnarray}
\phi(x,t;\kappa) & = & cn^{-1}\big[ tanh(\pm\sigma\xi);\kappa  \big];\label{gensol2}\\
&\! = \! & K(\kappa) -\frac{z}{\sqrt{1-k}}+\frac{2\kappa-1}{6(1-\kappa)^{3/2}}z^3+\frac{-8\kappa^2+8\kappa-3}{40(1-\kappa)^{5/2}}z^5+....;\quad z\equiv  tanh(\pm\sigma\xi);\nonumber\\
\label{expantanh2}
\end{eqnarray}
now the Taylor expansion in $z$ will be singular at the self-dual value $\kappa=1/2$, since the powers $3,7,11,..$, etc, in $z$, vanish at such a point, and
only the terms of orders 1,5,9, etc, will be present;
this expansion must be compared directly with the previous case described in Eqs. (\ref{exptanh1}). Furthermore, the full expression is divergent at a second self-dual point, $\kappa=1$, including the zero-order term given by the elliptic integral $K(\kappa)$.
The profile of this solution looks like a kink-like soliton for $\kappa<1$ (see figure \ref{solatmin}),and it will take the vev at two vacua as boundary conditions (see figure \ref{mmsn} below, and Eq. (\ref{bc}) above for a similar description for the previous case).
On the other hand, although the solution for $\kappa>1$ may have an interesting  profile, the topology for the potential in this interval does not allow to impose appropriate boundary conditions at the minima, since they correspond to an infinite collection of disconnected
vacua.

\subsubsection{Expansion around the minima: a kink-like soliton will decay to a kink-like soliton}
\label{kinkkink}

In similarity with the expansion (\ref{aroundmax}), the expansion around minima for the potential $V_{2}$ will read
\begin{eqnarray}
V_{2}(mim+\Delta\phi;\kappa)=V_{2}(\Delta\phi;\kappa);
\label{aroundmin}
\end{eqnarray}
and hence there is not a functional change for the potential for the field fluctuation $\Delta\phi$ with respect to the original field $\phi$. Hence, the expression (\ref{gensol2}) can be considered as the solution for the field fluctuation $\Delta\phi$ around the minima; the general profile for this soliton looks like a kink solution (described in the figure (\ref{solatmin}) above) in the interval $\kappa<1$. 
Furthermore, the approximation of the expression (\ref{gensol2}) in the form $cn(\Delta\phi;\kappa)\approx 1-\frac{1}{2}(\Delta\phi)^2=tanh(\pm\sigma\xi)$, valid in the expansion around the minima, will lead also to kink-like solution for the quadratic approximation in $\Delta\phi$.
Therefore, if this kink-like solution (at a maximum) will decay to its vacuum configuration (at a minimum) described previously, then it will decay to a kink-like configuration.

In the SSB scenario described in the section \ref{kinkcusp} an arbitrarily light particle may decay 
to a particle with a mass that is independent on the modulus $\kappa$. For the case at hand,  an arbitrarily massive term at the limit $\kappa\rightarrow 1^{-}$ (Eq. (\ref{expan4})) will generate a mass independent on the modulus (Eq. (\ref{expan5})).

\subsection{$\kappa>1$}
\label{kappamay1}

\begin{toexclude}
From the beginning the value $\kappa=1$ has shown to be special; first, the potential $V_{1}$ is trivial according to the Eq. (\ref{potlimit}); however,
the signs of the derivatives in the expressions (\ref{der}), depend sensitively on the limits from the left and from the right, $\kappa\rightarrow 1^\pm$. Specifically the second derivative expanded around  $\kappa=1$ will read
\begin{eqnarray}
\frac{d^2V_{1}}{d^2 \phi}\approx 2m^2 (\kappa-1)\big(1+2sinh^2(\phi) \big)+\mathcal{O}(\kappa-1);
\label{sde}
\end{eqnarray}
thus $\frac{d^2V_{1}}{d^2 \phi}>0$ as $\kappa\rightarrow 1^+$, and $\frac{d^2V_{1}}{d^2 \phi}<0$ as $\kappa\rightarrow 1^-$, for arbitrary $\phi$. For example for $\phi=0$ we have a (relative) minimum as  $\kappa\rightarrow 1^+$, and a (relative) maximum as  $\kappa\rightarrow 1^-$ (see the figure \ref{mmcn}).
\end{toexclude}

\begin{figure}[H]
  \begin{center}
   \includegraphics[width=.55\textwidth]{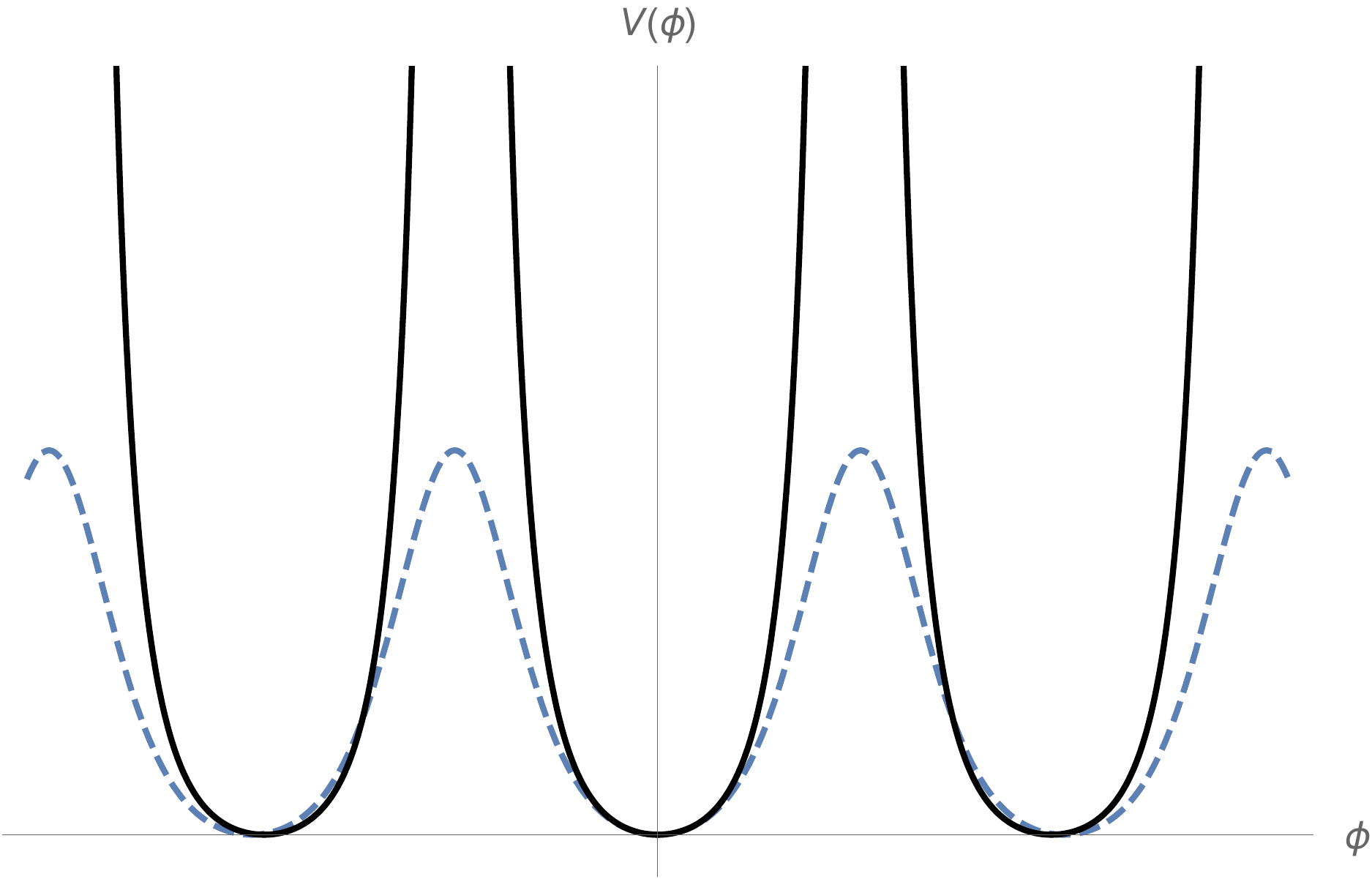}
  \caption{The potential $V_{2}$ for $\kappa>1$ is represented by the black curve; the dashed curve represents the same potential for $\kappa<1$.  This figure must be compared with $V_{1}$ described in the figure (\ref{mmcn}); an important difference is that, according to the Eq. (\ref{pot2}), the potential $V_{2}$ is not a constant just for $\kappa=1$.}  
  \label{mmsn}
\end{center}
\end{figure}

For the potential $V_{2}$, the differences between the profiles for $\kappa<1$, and for $\kappa>1$, are shown in the figure \ref{mmsn}; in relation to the change of topology, and  the topological defects associated with the geometry of the vacuum,
the conclusions are basically those shown in the figure (\ref{mmcn}) for $V_{1}$. However, as opposed to the  case described by the Eq. (\ref{sde}), the limits of the second derivative for $V_{2}$ as $\kappa\rightarrow 1^{\pm}$ coincide to each other (for example, the minima for both curves coincide to each other); note that the black curve diverges at the maxima of the dashed curve; as already commented, this fact will be clarified below by using the duality symmetry for the elliptic functions (see Eqs. (\ref{gauss3}), and (\ref{gauss4})).

\section{The continuous limits $\kappa\rightarrow 0$, and $\kappa\rightarrow 1$}
\label{conlimit}

The Lagrangian (\ref{intro1}) is related with the sine-Gordon model through an analytic continuation 
 \begin{eqnarray}
	b \rightarrow ib, \quad \mu \rightarrow -\mu, \quad L_{SG} = (\partial_{\mu} \phi)^2 + 2\mu \cos (b\phi); \label{sng}
 \end{eqnarray}
  however nontrivial issues are hidden behind thins simple transformation, since the sinh-Gordon model has one vacuum state, while the sin-Gordon has an infinite number of vacua, labelled by $\phi_{n} = 2 \phi {n}/{b}$; solitons and anti-solitons can then to interpolate between two vacua. In the approach at hand, such a topology change in the potentials is manifested around the critical self-dual point $\kappa=1$.
Moreover,
in the approach at hand, the sine-Gordon and sinh-Gordon models will be obtained as continuous limits for the modulus from our potentials constructed from elliptic functions; additionally the duality symmetry will be realized  as a classical symmetry by invoking the Jacobi dualities for the elliptic functions (section \ref{GLS}).

These continuous limits are obtained through the calculus on the the space of the modulus $\kappa$, the $\kappa$-space; in the section \ref{kspace} we describe briefly that calculus, which is encoded in the symbolic manipulation of mathematica.

\subsection{Double sin-Gordon model as the limit $\kappa\rightarrow 0$}
The double Sin-Gordon equation corresponds to a nonintegrable model of the form $\phi_{tt}-\phi_{xx}+\alpha\sin\phi+\sin 2\phi=0$, where $\sin \phi/2$ represents the fundamental harmonic, and $\sin \phi$ is the second harmonic; this model is widely used in physics and engineering, from condensed matter systems such as ferromagnetic materials, charge density waves, liquid crystal dynamics, nonlinear optics, $^3He$ spin waves, etc, to propagation of optical pulses \cite{pendula}. Many solutions have been constructed by using different techniques; see for example \cite{Shi}, and \cite{Min}.

In the approach at hand, the double sin-Gordon equation corresponds to the limit for a small elliptic modulus; from Eq. (\ref{em}) and from the expansion of the expression (\ref{der}) as $\kappa \to 0$, one obtains
\begin{eqnarray}
\frac{1}{c^2}\partial_{t}^{2}\phi-\partial_{x}^{2}\phi-m^2\sin( \phi)+m^2\frac{\kappa}{4}\sin (2 \phi)+\mathcal{O}(\kappa,\kappa^2)=0;
\label{sg}
\end{eqnarray}
note that the second harmonic appears as a first order correction for the zero-order first harmonic; higher order harmonics will appear at higher $\kappa$-orders. Similarly for the potential (\ref{pot2}) the co\-rres\-pon\-ding double sin-Gordon equation is basically the same above equation but with a change of sign for the harmonics.
\subsection{Double sinh-Gordon model as the limit $\kappa\rightarrow 1$}
Furthermore, the hyperbolic version can be obtained by expanding $\frac{dV_{}}{d\phi}$ around $\kappa \to 1$; for the potential (\ref{potlimit}) the double sinh-Gordon expression reads
\begin{eqnarray}
\frac{dV_{1}}{d\phi}=m^2 \left[(\kappa-1)\sinh 2  \phi +\frac{(\kappa-1)^2}{2}\sinh 4 \phi+ \mathcal{O}(\kappa, \kappa^2)\right];
\label{ng}
\end{eqnarray}
note that, due to the presence of the global factor $(\kappa-1)$ in the expression (\ref{der}), the first ''harmonic'' is generated at first order, and the second ''harmonic'' at the second order; however for the case (\ref{pot2}), we have
\begin{eqnarray}
\frac{dV_{2}}{d\phi}=m^2( \sinh 2 \phi+\frac{\kappa-1}{2}\sinh 4 \phi + \mathcal{O}(\kappa, \kappa^2));
\label{hg2}
\end{eqnarray}
thus, the second ''harmonic'' is generated as a first-order correction.

Exact travelling  waves solutions have been constructed by this case by using  different methods; see for example \cite{Waz,Abdul}.
Furthermore, the sinh-Gordon field theory has a wide range of physical applications, from a toy model for quantum gravity \cite{Lar}, to bosonic gas models \cite{kor, Bas}.
Moreover, the equilibrium and out-of-equilibrium properties in classical field theories have been studied by using sinh-Gordon models as prototypes of integrable field theories \cite{Mussa,Vecc}.
These scenarios will be explored from the perspective the our elliptic functions based formulation in future works.

\section{Jacobi duality symmetries}
\label{GLS}
As well known the elliptic functions are related to each other by a scaling of the variable and a transformation in the elliptic modulus,
\begin{eqnarray}
\kappa sn \left( z;\kappa\right)=sn\left(\kappa z; \frac{1}{\kappa}\right), \quad dn \left(z; \kappa \right)= cn\left(\kappa z; \frac{1}{\kappa}\right), \quad dn \left(\kappa z; \frac{1}{\kappa}\right)=cn(z;\kappa);
\label{landen1}
\end{eqnarray}
these transformations, due originally to Jacobi himself \cite{Jacobi},  induce the following transformations on our potentials,
\begin{eqnarray}
V_{1}=m^2\frac{cn^2( \phi, \kappa)}{dn^2( \phi, \kappa)}=m^2\frac{dn^2(\kappa \cdot  \phi, \frac{1}{\kappa})}{cn^2(\kappa \cdot  \phi, \frac{1}{\kappa})}; \label{landen2} \quad \\
V_{2}=m^2\frac{sn^2( \phi, \kappa)}{dn^2( \phi, \kappa)}=m^2\frac{1}{\kappa^2}\frac{sn^2(\kappa \cdot  \phi, \frac{1}{\kappa})}{cn^2(\kappa \cdot  \phi, \frac{1}{\kappa})}; \label{landen3} 
\end{eqnarray}
these transformations map a potential with small (large) modulus $\kappa$ to one of large (small) modulus; these potentials and their dual versions have been described in the figures (\ref{mmcn}), and (\ref{mmsn}), for certain fixed values for the modulus, for $\kappa<1$ and for $\kappa>1$; in particular the  periodic divergences in those figures can be identified now with the zeros of the $cn$ function in the above expressions. The limit $\kappa\rightarrow 1$ defines a critical value  for this dual mapping, and it will allow us to approach to the self-dual point for the Sinh-Gordon field theory discussed previously. Additionally the description for the potentials as functions on the field $\phi$, and on the modulus for the full range $\kappa\in R$, are given below in the figures.

This duality symmetry can be realized as a classical symmetry of the Lagrangian (\ref{lag}) under the following considerations;  the scaling on the field variable $ \phi^{\prime} \to \kappa\cdot  \phi$, implies that the kinetic term in the Lagrangian will be invariant provided that, the space-time coordinates scale isotropically as $x^{\prime} \to k x$, and $t^{\prime}\to k t$; hence $L( \phi, m;x,t;\kappa)=L( \phi^{\prime}, m; x^{\prime}, t^{\prime};\kappa^{\prime})$, with $\kappa^{\prime}\to 1/\kappa$. 

Let $V'$ be the versions with arguments $(\kappa \cdot  \phi, \frac{1}{\kappa})$ for the potentials $V_{1}$, and $V_{2}$; then the Eqs. (\ref{landen2}), and (\ref{landen3}) can be rewritten as
\begin {eqnarray} V_{1} &=&m^4 \frac{1}{V'_{1}}, \label {gauss1}\\
 V_{2}& =& \frac{m^2}{\kappa} \frac{V'_{2}}{ (1-\kappa) V'_{2}+\kappa m^2}; \label{gauss2} \end {eqnarray} 
the first equation is established in a direct way; however, the second one is more subtle, which is established by using the identities for elliptic functions with modulus $\kappa' = \frac{1}{\kappa}$.  Furthermore, these duality maps generate singularities, namely,
\begin{eqnarray}
V_{1}'=0\rightarrow \phi=(2n+1)\frac{K(1/\kappa)}{\kappa};\label{gauss3}\\
  (1-\kappa) V'_{2}+\kappa m^2=0\rightarrow 
\phi=(2n+1)\frac{K(1/\kappa)}{\kappa};
\label{gauss4}
\end{eqnarray}
note that, in spite of their appearances, the duality maps have the same singular points for the field $\phi$, and defined by the integral periods. These periodic singularities describe the singular potentials for $\kappa>1$ in the figures 
(\ref{mmcn}), and (\ref{mmsn}). Note also that just  at the so called  self-dual point, $\kappa=1$, the mapping (\ref{gauss2}) is trivial, and thus it is excluded in the expression (\ref{gauss4}).

Specifically near the self-dual point the expressions (\ref{gauss1}) and (\ref{gauss2}) imply that 
\begin{eqnarray}	
V_{1} \rightarrow \frac{1}{V_{1}},\quad   {\rm and} \quad  V_{2} \rightarrow V_{2}, \quad   {\rm as} \quad  \kappa \rightarrow 1; \label{sd2}
\end{eqnarray}
hence, in the second case the duality transformation is trivial; however in the first case this transformation is generating singularities, due basically to the zeros of the $cn$ function, according to the Eq. (\ref{gauss3}). In the figures below we describe the potentials and their dual versions as functions on the field $\phi$ and the modulus.

\begin{figure}[H]
  \begin{center}
   \includegraphics[width=.55\textwidth]{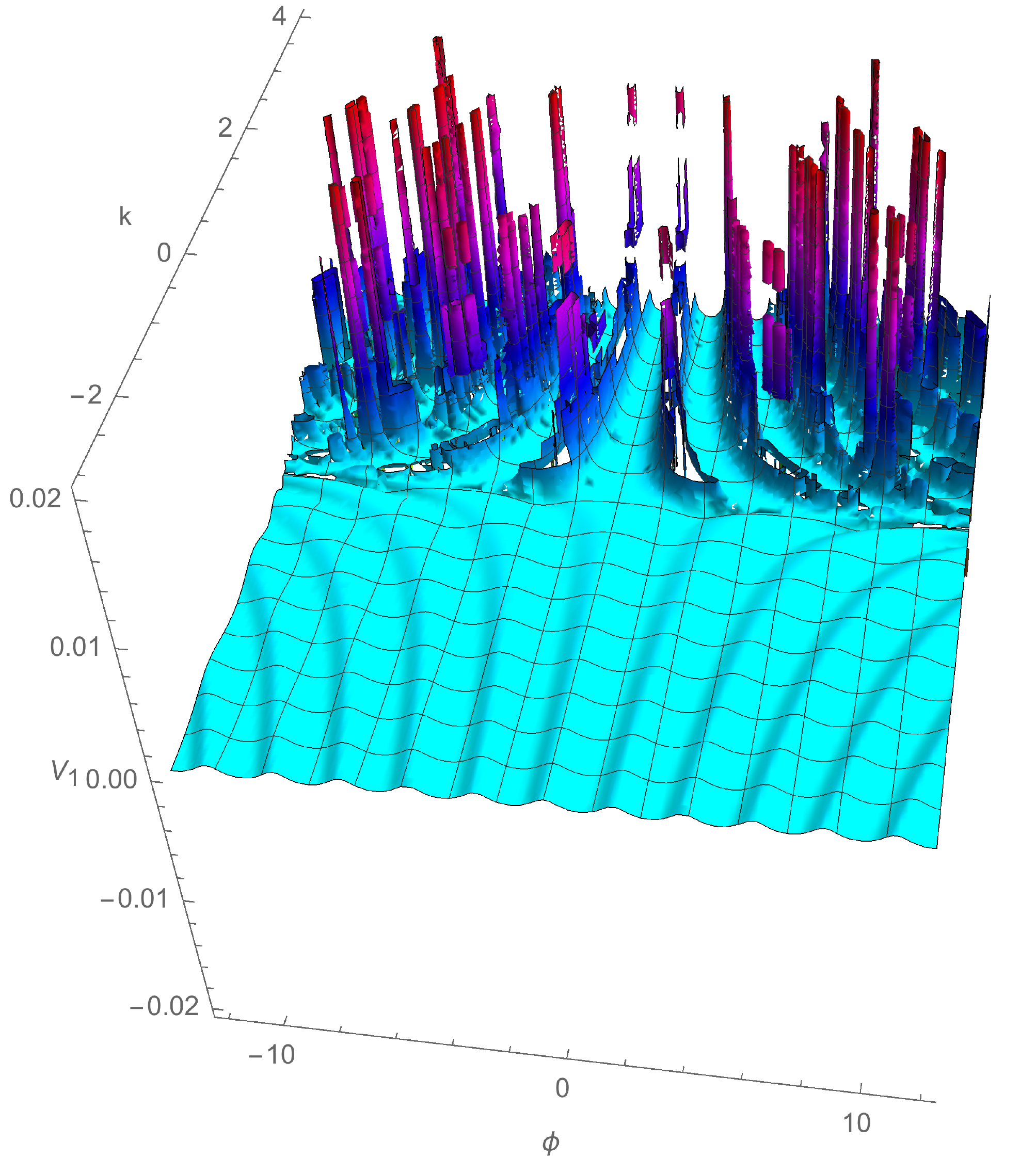}
  \caption{  A distant view for the potential $V_{1}$ as function on $\phi$, and on the $\kappa$-modulus; it shows the weak regime  $\kappa<1$ (including the negative values for the modulus), where the potential is finite, without singularities; for $\kappa>1$  the divergences are shown.}  
  \label{v1kx}
\end{center}
\end{figure}
\begin{figure}[H]
  \begin{center}
   \includegraphics[width=.55\textwidth]{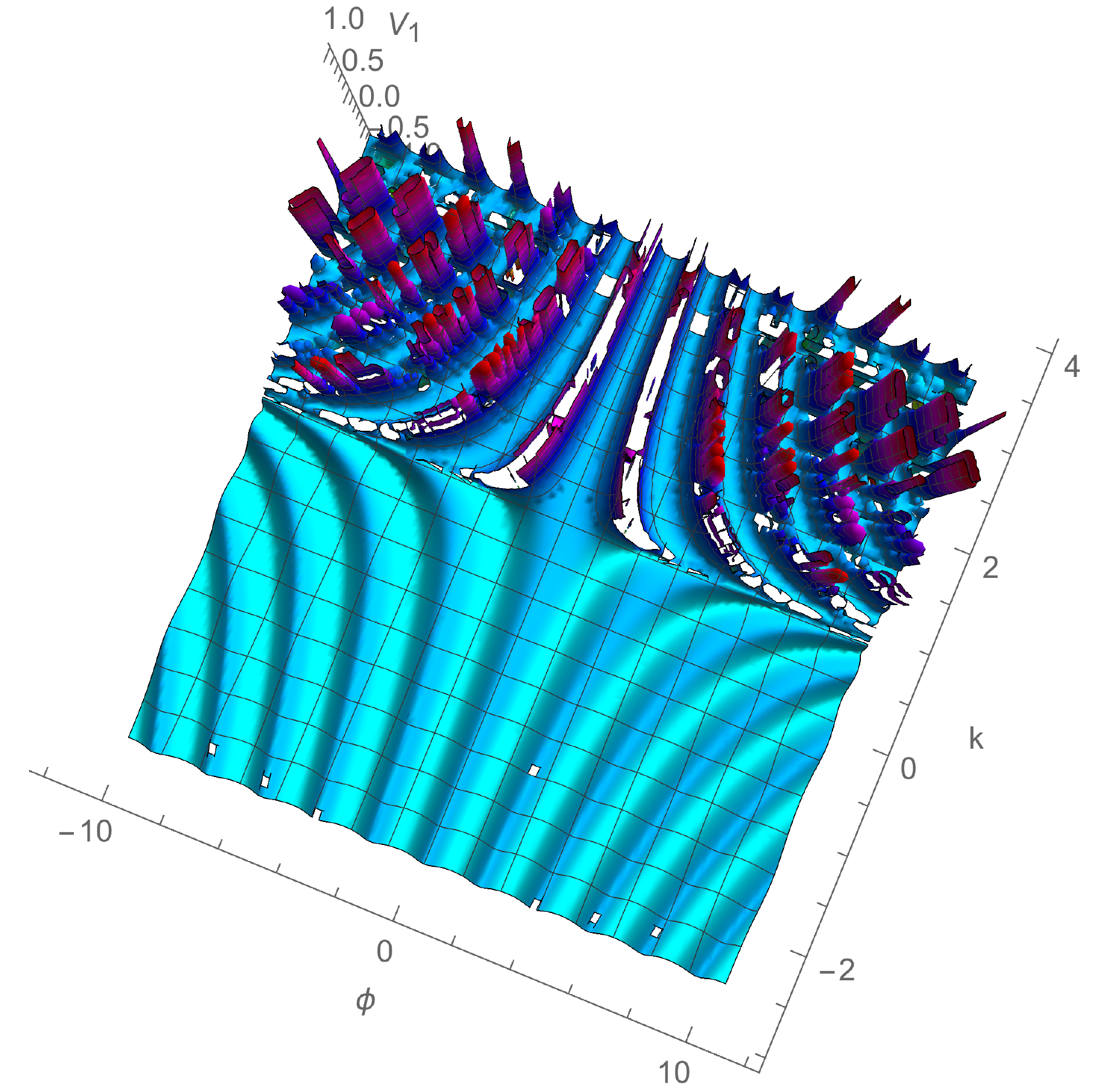}
  \caption{ View from above for $V_{1}$; it shows clearly the critical role of the self-dual point $\kappa=1$; the duality is mapping the potential with $\kappa<1$, to the potential for $\kappa>1$, with the appearance of the singularities. note the correspondence between the regions for the maxima for $\kappa<1$, and the regions for the minima for $\kappa>1$; the curves described in  the figure \ref{mmcn} correspond to two fixed values for the modulus, namely, for $\kappa<1$, and for $\kappa>1$.}  
  \label{v1kkxx}
\end{center}
\end{figure}

\begin{figure}[H]
  \begin{center}
   \includegraphics[width=.55\textwidth]{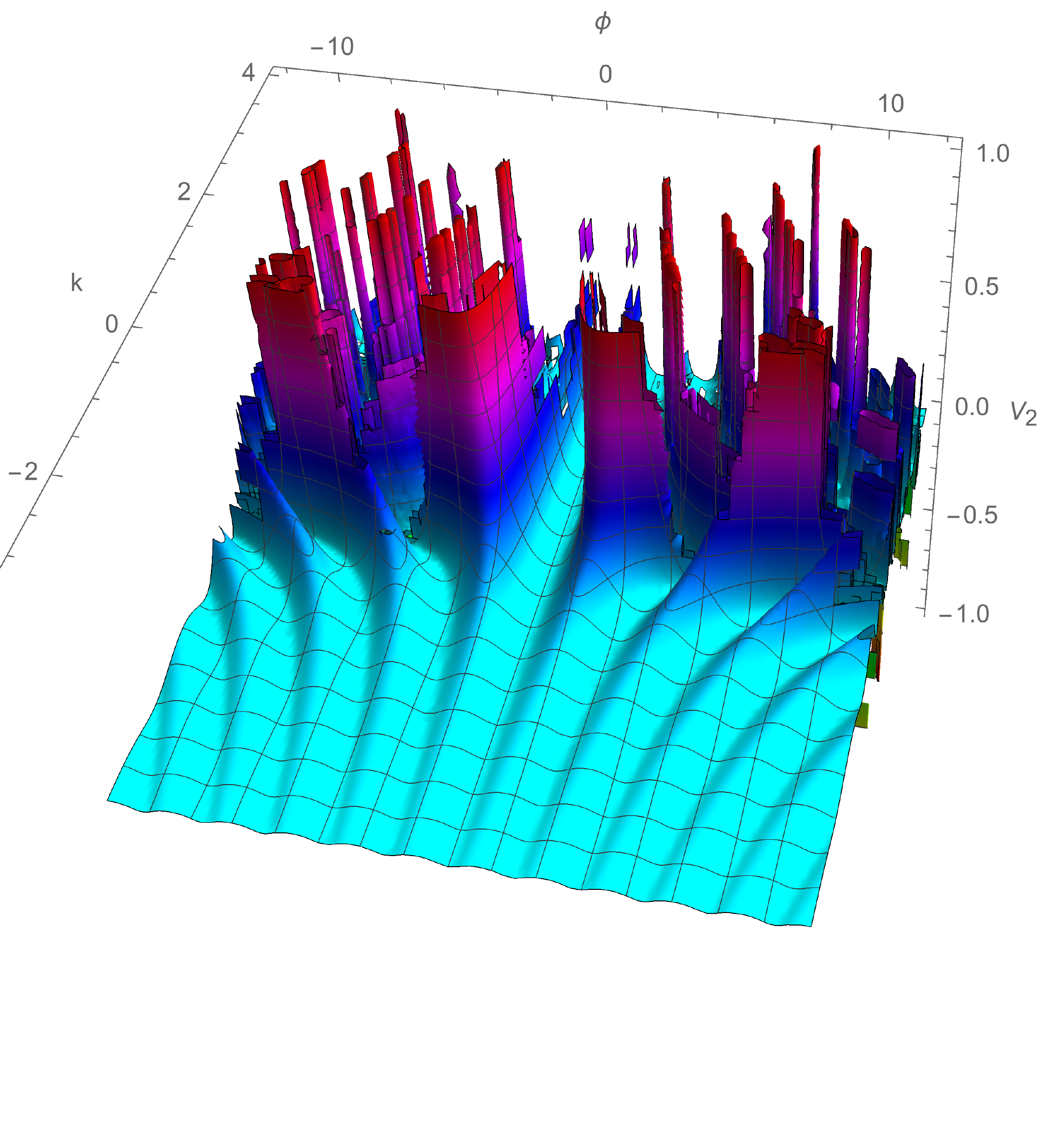}
  \caption{  A close view for the potential $V_{2}$; note the correspondence between the regions for the minima for $\kappa<1$, and the regions for the minima for $\kappa>1$; the curves described in  the figure \ref{mmsn} correspond to two fixed values for the modulus, namely, for $\kappa<1$, and for $\kappa>1$.}  
  \label{v2kx}
\end{center}
\end{figure}
\begin{figure}[H]
  \begin{center}
   \includegraphics[width=.55\textwidth]{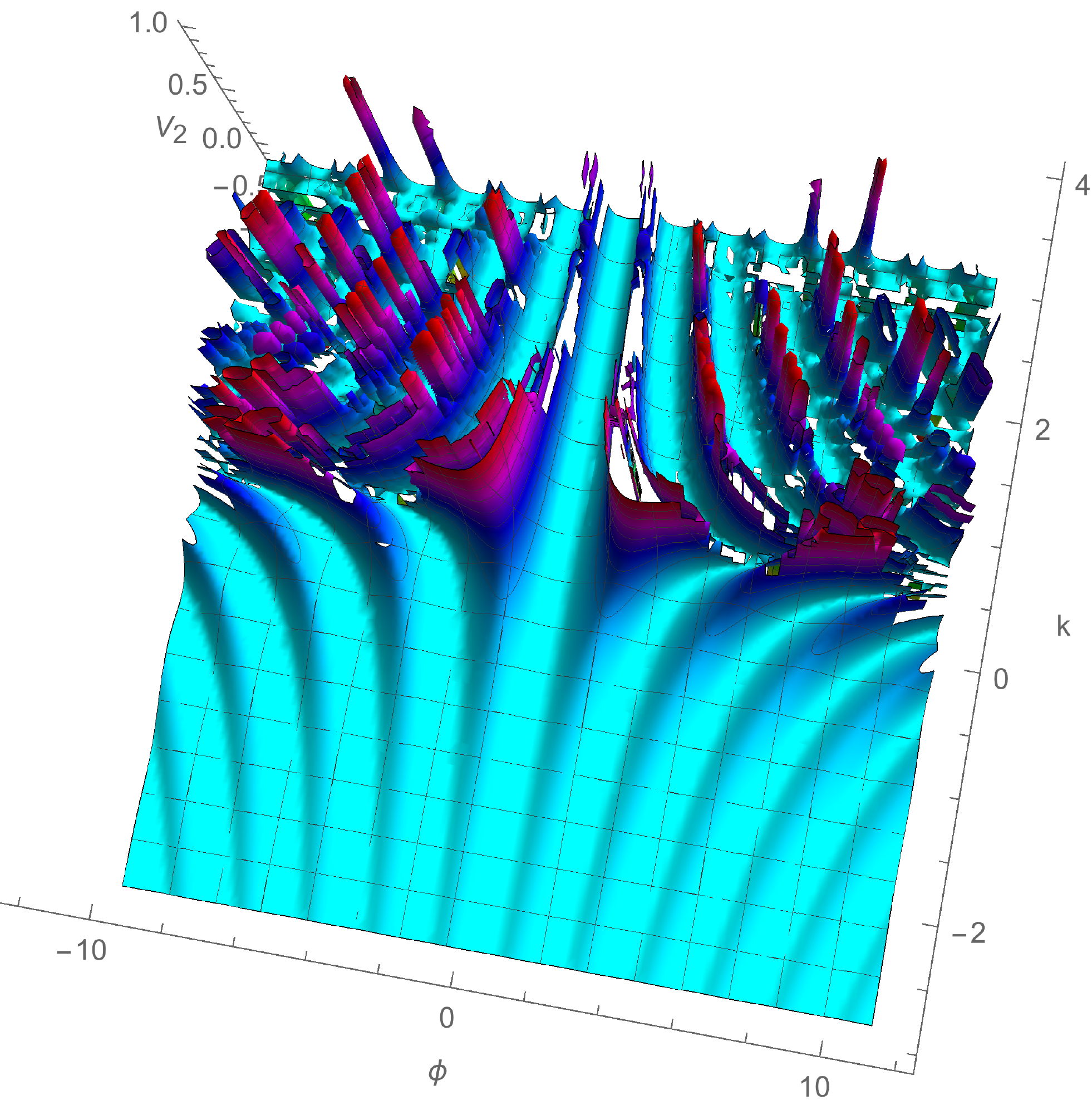}
  \caption{  A view from above for the potential $V_{2}$; note the correspondence between the regions for the minima for $\kappa<1$, and the regions for the minima for $\kappa>1$; the curves described in  the figure \ref{mmsn} correspond to two fixed values for the modulus, namely, for $\kappa<1$, and for $\kappa>1$.}  
  \label{v22kx}
\end{center}
\end{figure}
\begin{figure}[H]
  \begin{center}
   \includegraphics[width=.55\textwidth]{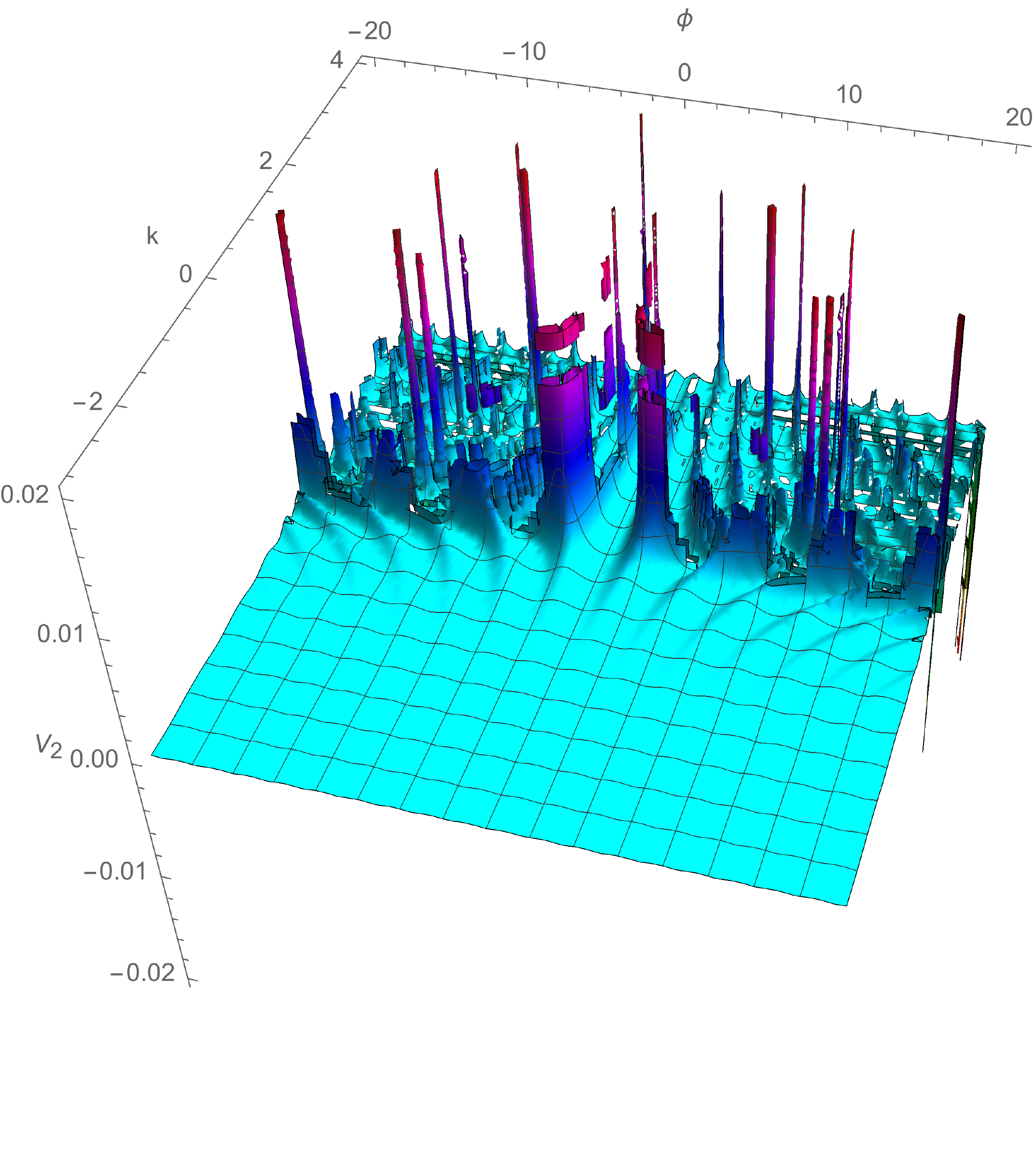}
  \caption{  A distant view for the potential $V_{2}$; note that at the limit  $\kappa\rightarrow -\infty$ the potential is asymptotically flat.}  
  \label{v222kx}
\end{center}
\end{figure}

We consider now the effect of the duality transformations on the local expressions for the potentials around the maxima and minima; first, the expressions (\ref{aroundmax}), and (\ref{aroundmin}) around the maxima and the minima for the potentials $V_{1}$ and $V_{2}$  do not change respect to the original expressions; thus the above figures describe also locally the respective potentials. However, if we consider now the expressions (\ref{minex}), and (\ref{maxex}), then they are transformed as
\begin{eqnarray}
V_{1}(min+\Delta\phi; \kappa)=m^2 sn^2(\Delta\phi;\kappa)=\frac{m^2}{\kappa^2}sn^2(\kappa\Delta\phi;1/\kappa)=\frac{1}{\kappa^2}V'_{1};\label{local1}\\
V_{2}(max+\Delta\phi;\kappa)=m^2\frac{cn^2(\Delta\phi;\kappa)}{1-\kappa}=m^2\frac{dn^2(\Delta\phi;\kappa)}{1-\kappa}=m^2-V'_{2};
\label{local2}
\end{eqnarray}
according to the transformations (\ref{landen1}); these potentials are illustrated in the figures below.
\begin{figure}[H]
  \begin{center}
   \includegraphics[width=.55\textwidth]{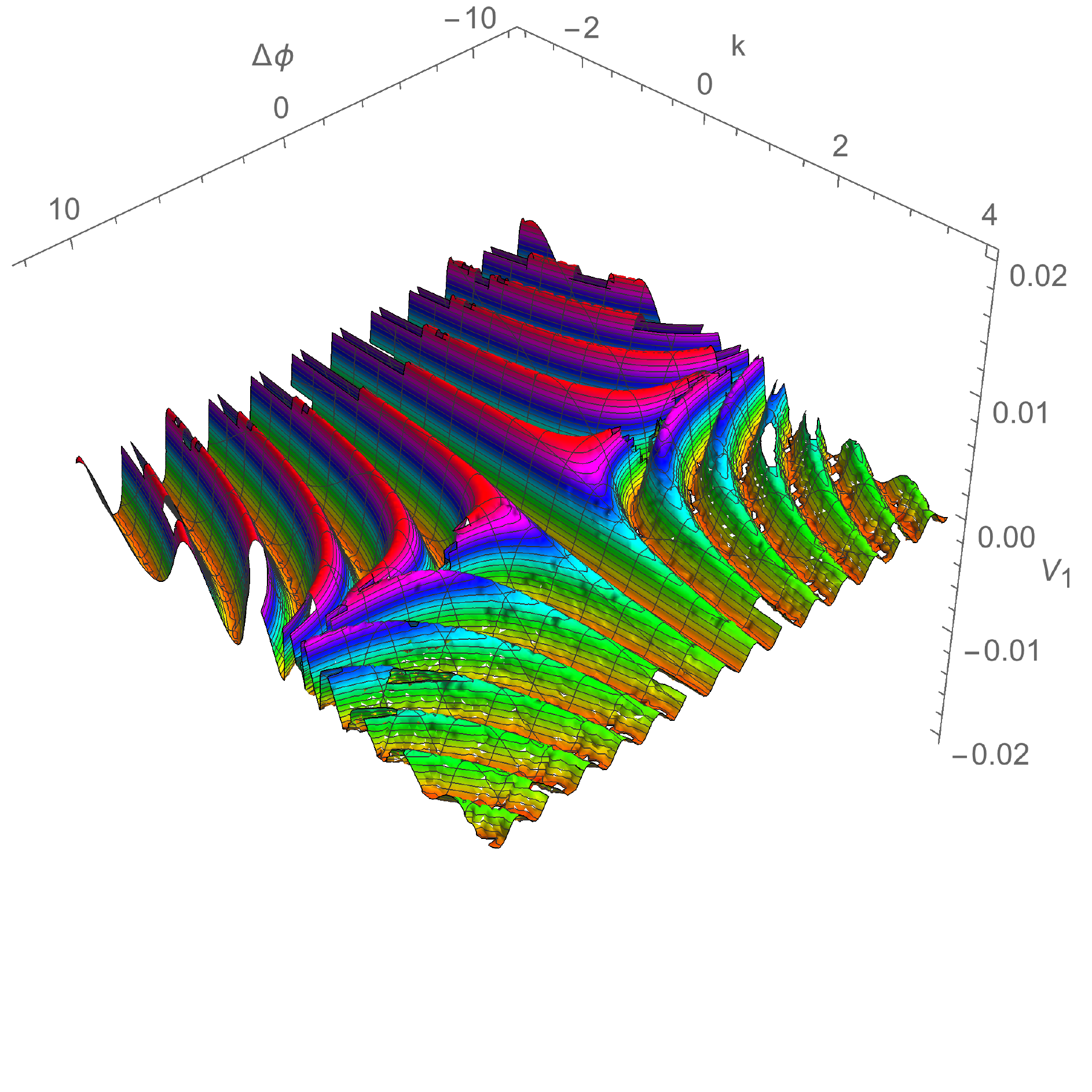}
  \caption{  The potential around the minima of $V_{1}$ is finite for all the range $\kappa\in R$; the green zone corresponds to $\kappa<1$; the potential for $\kappa>1$ is obtained by the duality and corresponds to the red zone. Note that in particular there are not divergences for the range $\kappa>1$, as opposed to the figure \ref{v1kx}.}  
  \label{v1local}
\end{center}
\end{figure}
\begin{figure}[H]
  \begin{center}
   \includegraphics[width=.55\textwidth]{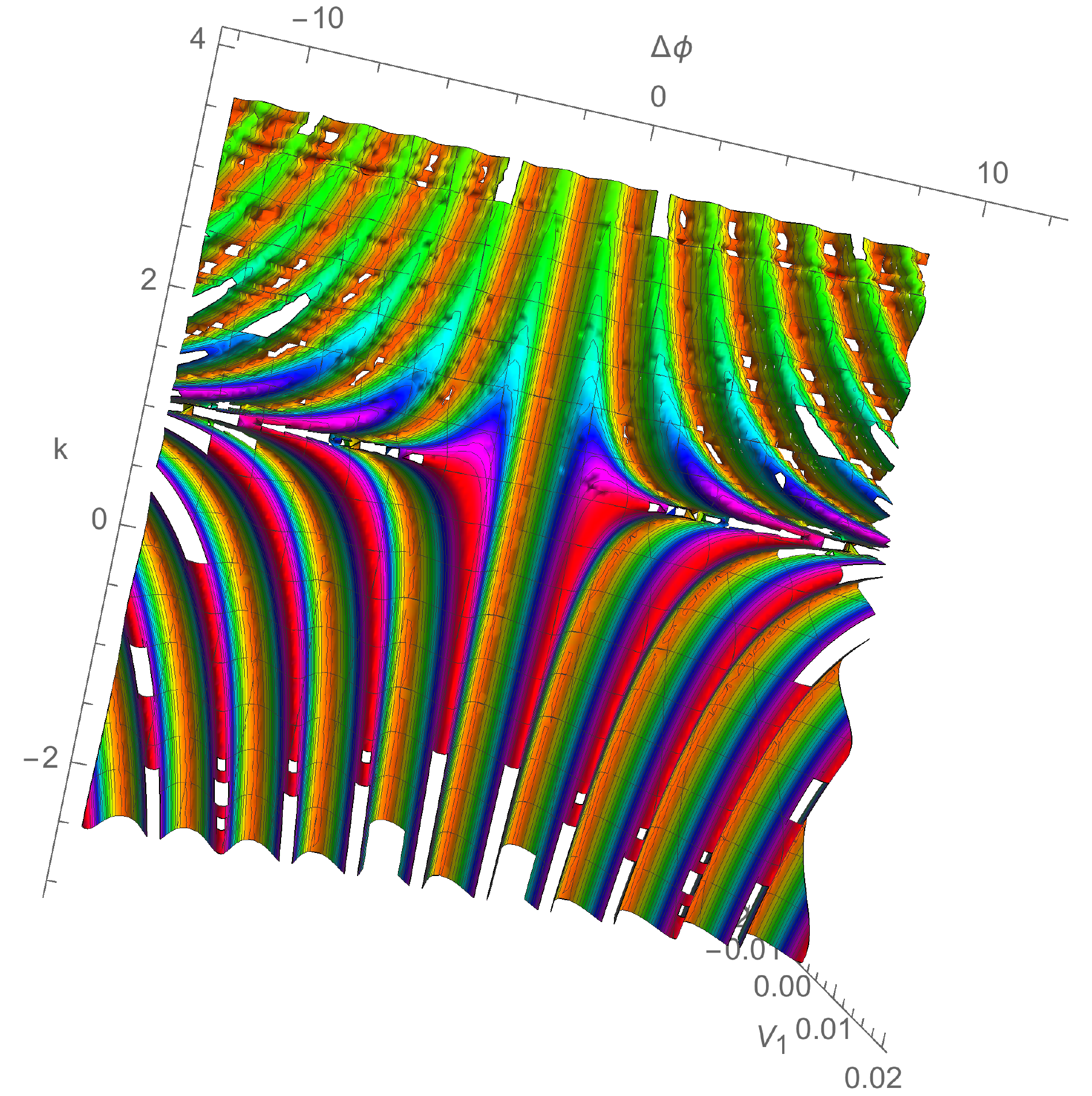}
  \caption{A view from above; the potential around the minima of $V_{1}$, although finite, it is not continuous at $\kappa\rightarrow 1$.}  
  \label{v11local}
\end{center}
\end{figure} 

\begin{figure}[H]
  \begin{center}
   \includegraphics[width=.55\textwidth]{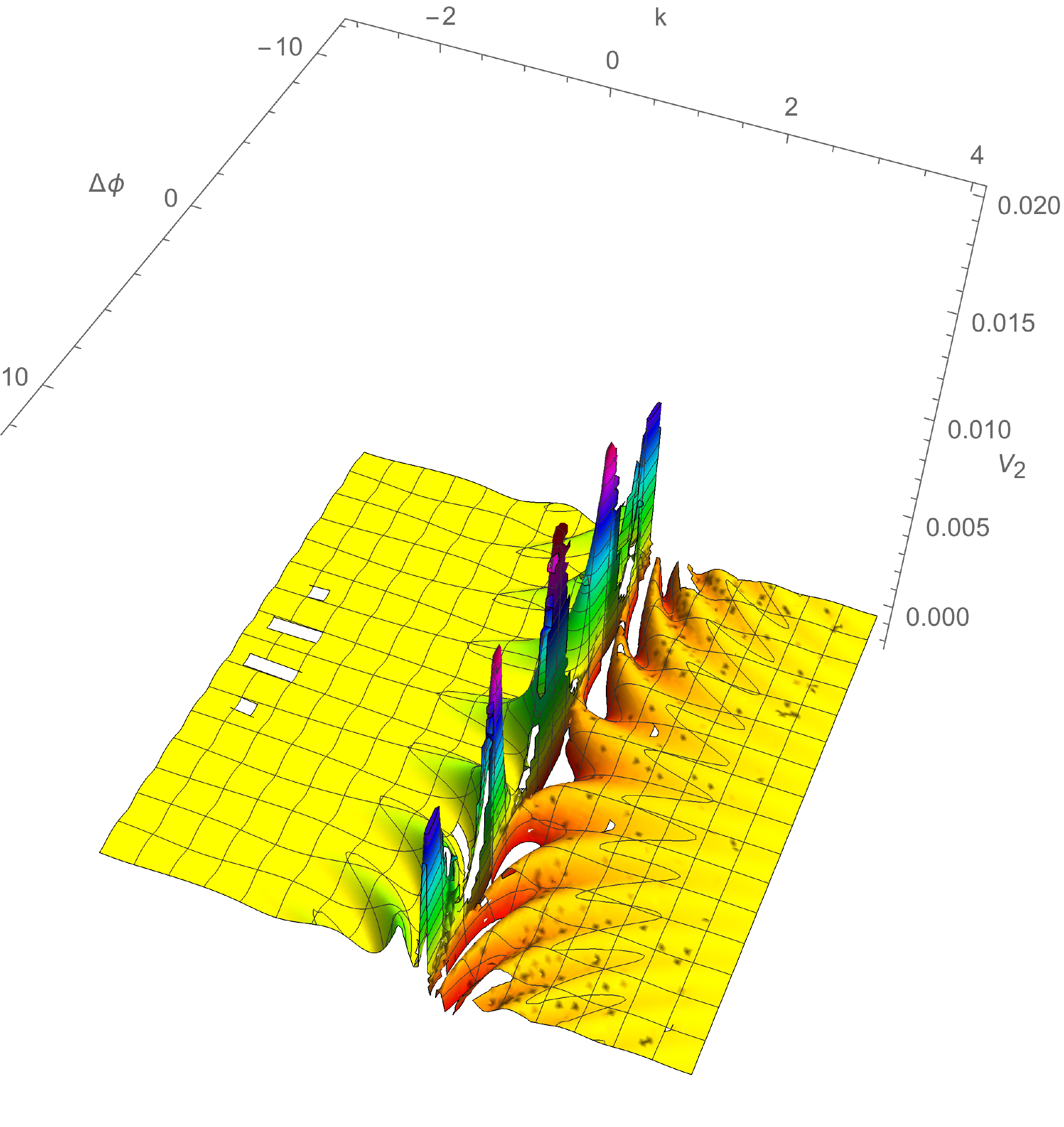}
  \caption{ A distant view for the potential around the maxima for $V_{2}$; the potential diverges as $\lim_{_{\kappa\rightarrow 1^{\pm}}} V_{2}=\mp\infty$; as opposed to the figure \ref{v222kx}, the singularities are located only at $\kappa=1$.}  
  \label{v2local}
\end{center}
\end{figure} 

\subsection{Approaching to the self-dual point: $V_{1}$}

\begin{figure}[H]
  \begin{center}
   \includegraphics[width=.55\textwidth]{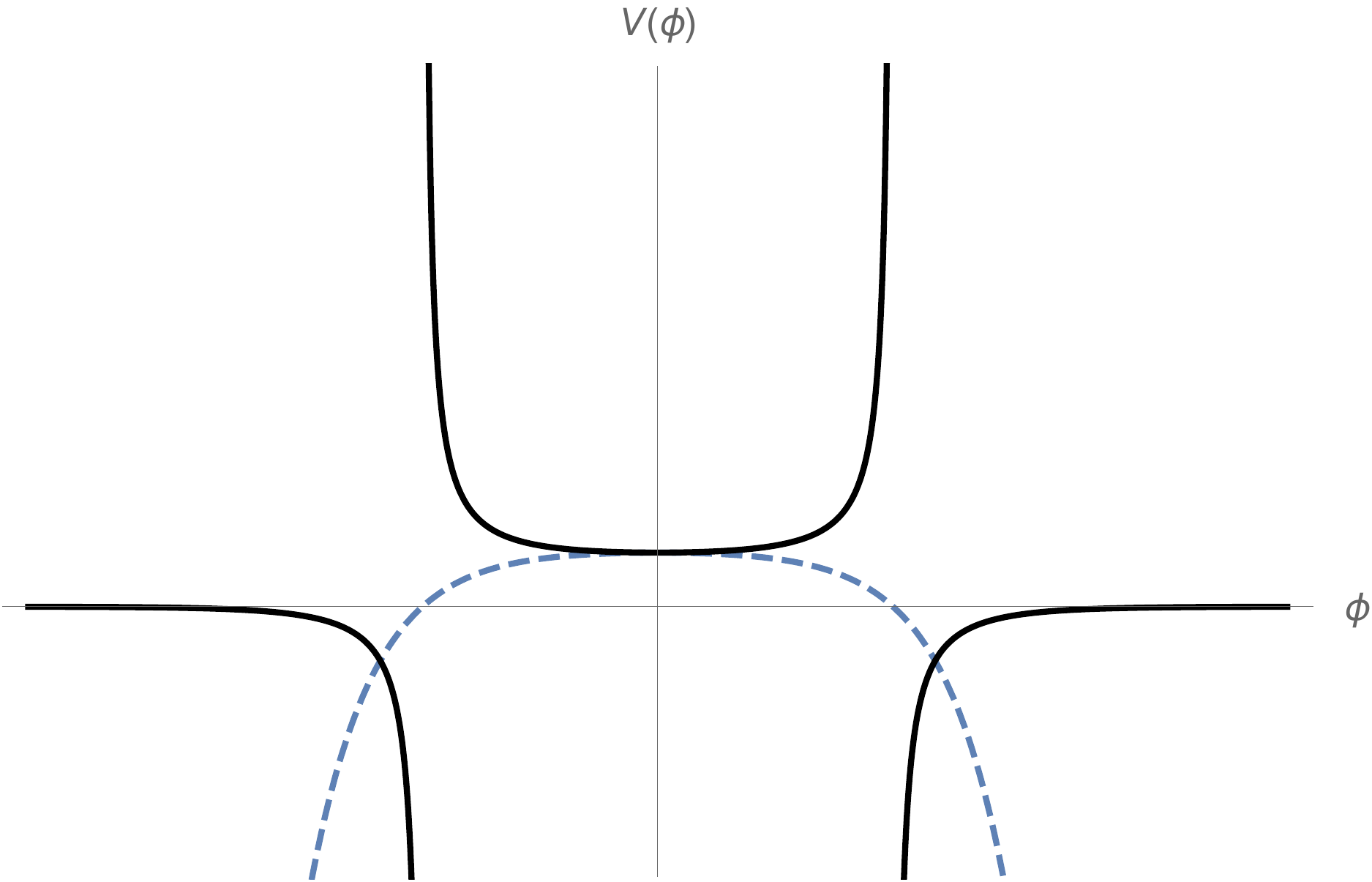}
  \caption{$V_{1}$: The approaching $\kappa\rightarrow 1$ for $\kappa<1$; the discontinuities in the black curve correspond to the two real roots in Eq. (\ref{mapp2}).}. 
  \label{selfdual1}
\end{center}
\end{figure}

\begin{figure}[H]
  \begin{center}
   \includegraphics[width=.55\textwidth]{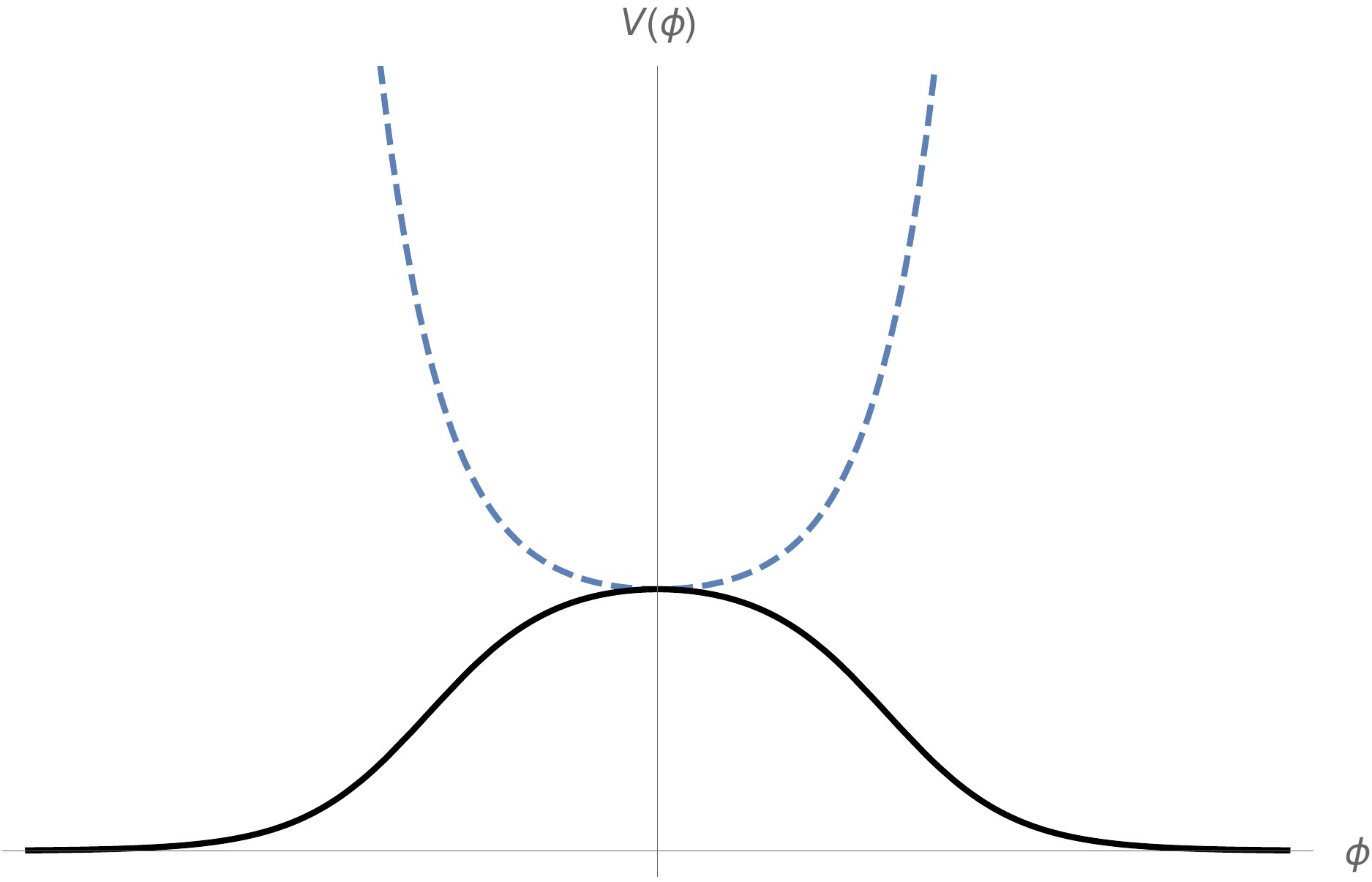}
  \caption{The approaching $\kappa\rightarrow 1$, for $\kappa>1$.}  
  \label{selfdual2}
\end{center}
\end{figure}

The expansion of the potential $V_{1}$ around $\kappa = 1$ will read
 \begin{eqnarray}
	\frac{1}{m^2}V_{1}(\phi,\kappa) \approx 1 + (\kappa -1) \sinh^2 (\phi) +\mathcal{O}(\kappa-1), \quad {\rm as} \quad \kappa \rightarrow 1; 
\label{mapp1}
 \end{eqnarray}
for the limit $ \kappa \rightarrow 1$ from the left, it means $\kappa < 1$, this potential is shown by the dashed curve in the figure (\ref{selfdual1}); its inversion under the duality (\ref{sd2}) is shown by the (disconnected) black curve, with singularities at the zeros of the dashed curve,  namely
\begin{eqnarray}
	\sinh  \phi = \pm \sqrt{\frac{1}{1-\kappa}}; 
\label{mapp2}
 \end{eqnarray}
Similarly for the limit $\kappa \rightarrow 1$ from the right it means $\kappa > 1$, the potential  $V_{1}$ is shown in the figure (\ref{selfdual2}) by the dashed curve; its inversion under duality is described by the black curve, without singularities, since for $\kappa>1$ there not exist real solutions for  Eq. (\ref{mapp2}).
\subsection{ $V_{2}$: a self-dual potential at the self-dual point}
\begin{figure}[H]
  \begin{center}
   \includegraphics[width=.55\textwidth]{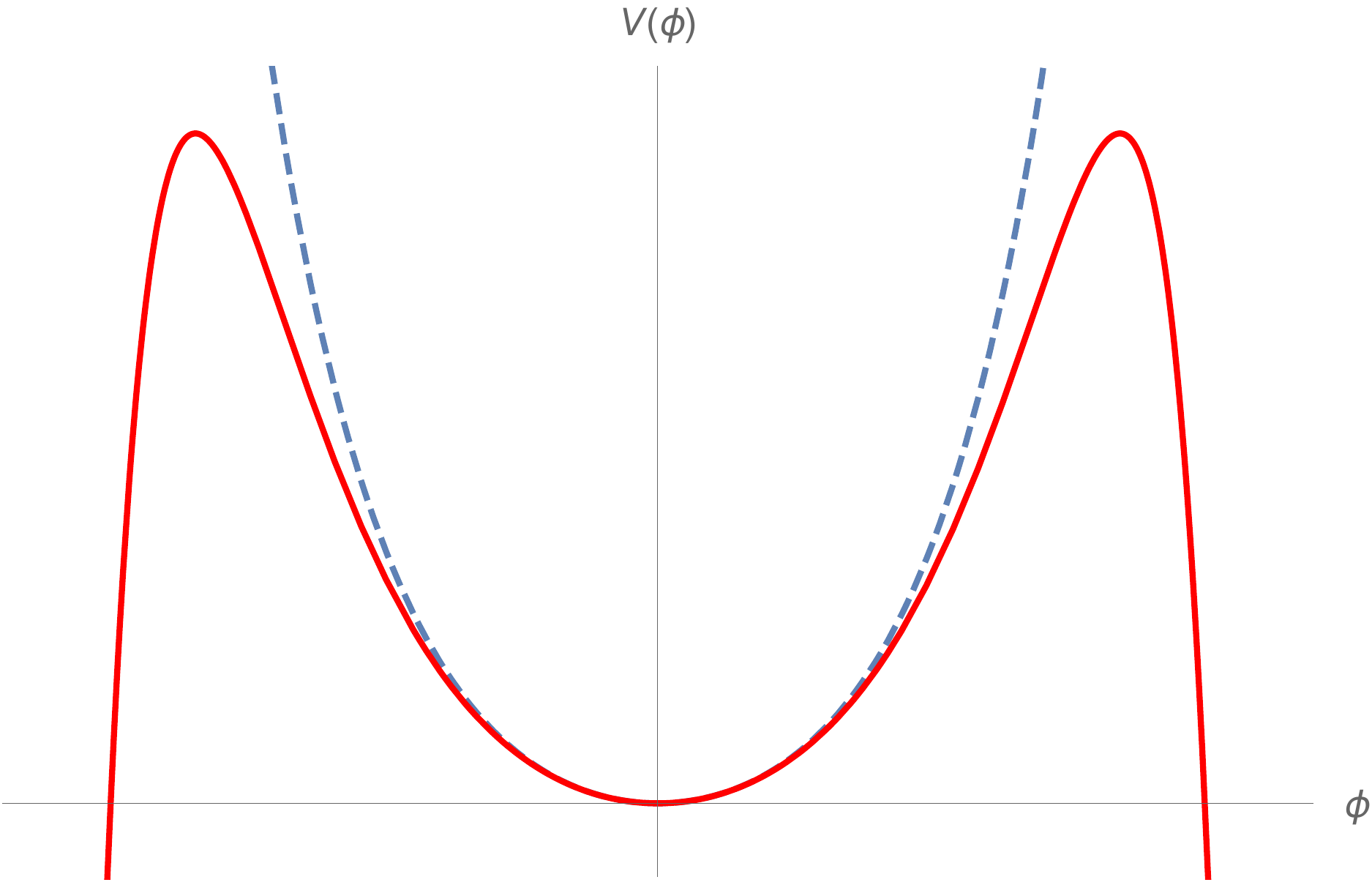}
  \caption{The approaching $\kappa\rightarrow 1^+$ represented by the dashed curve, with a absolute minimum at $\phi=0$; the red curve represents the limit  $\kappa\rightarrow 1^-$, with a relative minimum at $\phi=0$, and additionally the appearance of two maxima. In this case, as opposed to the figures (\ref{selfdual1}), and (\ref{selfdual2}), the black curves coincide with these curves, since the potential $V_{2}$ is self-dual at the self-dual point according to the expression (\ref{sd2}). }  
  \label{selfdual33}
\end{center}
\end{figure}

Similarly the expansion of the potential $V_{2}$  around $\kappa = 1$ will read 
\begin{eqnarray}
\frac{1}{m^2}V_{2} ( \phi, \kappa) \approx \sinh^{2} ( \phi) + \frac{\kappa - 1 }{8} \sinh ( \phi) [\ - 7 \sinh ( \phi) + \sinh (3  \phi) + 4  \phi \cdot \cosh ( \phi) ]\ +  \mathcal{O}(\kappa-1)
\label{mapp3};
\end{eqnarray}
this approach to the self-dual point is shown in the figure (\ref{selfdual33}).

\section{An approach from the $\kappa$-space perspective}
\label{kspace}
In the approach at hand we have used the calculus for the elliptic functions related to their main argument, the field $\phi$.  However, the calculus respect to the $\kappa$-modulus is also available; in fact, the continuous limits described in the section (\ref{conlimit}) are obtained through the Taylor expansions of the elliptic functions in the $\kappa$-space; specifically the right hand side of the expression (\ref{der}) are expanded in Taylor series in order to obtain the expression (\ref{ng}). Moreover, such a calculus on the $\kappa$-space is not closed respect to the three fundamental elliptic functions, since it requires to invoke also the elliptic integrals. We give here, as example of a closed expression within the elliptic functions/integrals formalism, the derivative of the potential $V_{1}$ in $\kappa$-space;
\begin{eqnarray}
\frac{d}{d\kappa}V_{1}=\frac{cn(\phi,\kappa)sn(\phi,\kappa)\Big(E(Am(\phi,\kappa),\kappa)-\kappa cd(\phi,\kappa)sn(\phi,\kappa)+(\kappa-1)\phi\Big)}{\kappa(1-\kappa)dn(\phi,\kappa)}\nonumber\\
-\frac{cn^3(\phi,\kappa)sn(\phi,\kappa)\Big(E(Am(\phi,\kappa),\kappa)- dn(\phi,\kappa)sc(\phi,\kappa)+(\kappa-1)\phi\Big)}{\kappa(1-\kappa)dn^3(\phi,\kappa)};\end{eqnarray}
where $cd(\phi,\kappa)\equiv\frac{cn(\phi,\kappa)}{dn(\phi,\kappa)}$,  $sc(\phi,\kappa)\equiv\frac{sn(\phi,\kappa)}{cn(\phi,\kappa)}$, $Am(\phi,\kappa)$ stands for the Jacobi amplitude, which appears as the argument for the elliptic integral of the second kind, $E$; note the critical role of the self-dual points $\kappa=0$, and $\kappa=1$.
This calculus 
will allow us to explore the strong/weak dualities in the $\kappa$-space, for any model based on elliptic functions.

\section{Concluding remarks} 

In this work we have focused our attention only on the specific Jacobi transformation closer to the standard sin/sinh
duality symmetry manifested at the level to the S-matrix; there are some obvious questions that it will be interesting to explore, for example to determine the effect of the other Jacobi transformations on the elliptic functions based potentials constructed here, and to study the different regimes that define in the $\kappa$-space.

The standard sin/sinh Gordon field theory manifests the strong/weak duality at the S-matrix level but not at classical level. Conversely our elliptic functions based formulation manifests such a symmetry at classical level, and it is mandatory the construction of the corresponding S-matrix, in order to determine explicitly its dependence on the $\kappa$-modulus; works along these lines are in progress. It would be interesting to provide a possible holographic interpretation 
for the $\kappa$-modulus, con\-si\-dering that for the classical pendulum the $\kappa$-modulus is identified directly with the energy of the system.\\

{\bf Acknowledgements}
This work was supported by the Sistema Nacional de Investigadores (SNI, Mexico), and the Vicerrector\'ia de Investigaci\'on y Estudios de Posgrado (VIEP-BUAP). K. Peralta-Martinez and D. A.  Zarate-Herrada would like to thank Consejo Nacional de Ciencia y Tecnolog\'ia (CONACYT, Mexico) for financial support (CVU: 1008062 and 942384 respectively). Graphics and the symbolic manipulation of elliptic functions have been made using Mathematica.

\end{document}